\documentclass[sigconf,table]{acmart}

\newif\ifshowTODOs
\showTODOsfalse

\usepackage{xcolor} %
\newcommand{\xxnote}[3]{}
\ifx\hidenotes\undefined
  \usepackage{color}
  \renewcommand{\xxnote}[3]{\color{#2}{\textbf{#1:} #3}}
\fi

\newcommand{\user}[1]{\textsf{\textit{\textcolor{black}{``#1''}}}}

\AtBeginDocument{%
  }

\renewcommand\footnotetextcopyrightpermission[1]{}

\usepackage{subcaption}
\usepackage{siunitx}
\usepackage{multirow}
\usepackage{comment}
\usepackage{xspace}
\usepackage{hyperref}

\usepackage[onecolumn]{structuralmap}   %
\usepackage{graphicx}
\usepackage{booktabs}
\usepackage{array}
\usepackage{tabularx}
\usepackage{xcolor}
\usepackage{colortbl}
\showmapfalse   %

\definecolor{ICONRevTQBAU}{HTML}{EAF3FC}
\definecolor{ICONRevTQICON}{HTML}{CDE2F7}
\definecolor{ICONRevPDBAU}{HTML}{EEF6E9}
\definecolor{ICONRevPDICON}{HTML}{D8EBCD}
\definecolor{ICONRevPWBAU}{HTML}{FFF2E5}
\definecolor{ICONRevPWICON}{HTML}{F9DEC1}
\definecolor{ICONRevMCBAU}{HTML}{F5EDFA}
\definecolor{ICONRevMCICON}{HTML}{E6D5F0}
\definecolor{ICONRevHeader}{HTML}{EEEEEE}
\definecolor{ICONRevGrid}{HTML}{D5D5D5}
\definecolor{ICONRevChanged}{HTML}{EAF0F5}

\newcommand{\ICONRevTQ}[1]{%
  \includegraphics[trim=56.69bp 1411.65bp 53.86bp 1451.34bp,clip,width=#1]{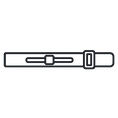}}
\newcommand{\ICONRevPD}[1]{%
  \includegraphics[trim=189.92bp 1000.63bp 181.42bp 941.10bp,clip,width=#1]{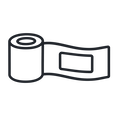}}
\newcommand{\ICONRevPW}[1]{%
  \includegraphics[trim=66bp 315bp 66bp 320bp,clip,width=#1]{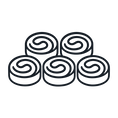}}
\newcommand{\ICONRevMC}[1]{%
  \includegraphics[trim=214bp 173bp 107bp 102bp,clip,width=#1]{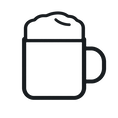}}

\newcommand{\ICONRevBodyWidthTQ}{31.2pt}
\newcommand{\ICONRevBodyWidthPD}{24pt}
\newcommand{\ICONRevBodyWidthPW}{24pt}
\newcommand{\ICONRevBodyWidthMC}{20pt}

\newcommand{\ICONRevTrialCell}[2]{%
  \cellcolor{ICONRev#1#2}%
  \begin{minipage}[c][43pt][c]{\linewidth}
    \centering
    \parbox[c][25pt][c]{\linewidth}{\centering%
      \csname ICONRev#1\endcsname{\csname ICONRevBodyWidth#1\endcsname}}\par
    \vspace{2pt}{\sffamily\bfseries\small #2}
  \end{minipage}}
\newcommand{\ICONRevSwatches}[1]{%
  {\setlength{\fboxsep}{0pt}%
   \colorbox{ICONRev#1BAU}{\rule{0pt}{6pt}\hspace{6pt}}\hspace{1pt}%
   \colorbox{ICONRev#1ICON}{\rule{0pt}{6pt}\hspace{6pt}}}}
\newcommand{\ICONRevLegendItem}[2]{%
  \mbox{\ICONRevSwatches{#1}\hspace{4pt}%
  \raisebox{-.2\height}{\csname ICONRev#1\endcsname{\csname ICONRevLegendWidth#1\endcsname}}\hspace{4pt}#2}}
\newcommand{\ICONRevChangedCell}{\cellcolor{ICONRevChanged}\textbf{Changed}}

\begin{document}

\title[Substantial, Decomposable, and Invisible: Visual Context Misalignment in Instructional Videos for Physical Tasks]{Substantial, Decomposable, and Invisible: Visual Context Misalignment in Instructional Videos for Physical Tasks}

\author{Yayuan Li$^{1,*}$\quad Chenglin Li$^{1,*}$\quad Jingying Wang$^{1}$\quad Filippos Bellos$^{1}$\quad \\ Anhong Guo$^{1,\dagger}$\quad Jason J. Corso$^{1,2,\dagger}$}
\affiliation{%
  \institution{$^{1}$University of Michigan \quad $^{2}$Voxel51}
  \city{}
  \country{}
}

\newcommand\blfootnote[1]{%
  \begingroup
  \renewcommand\thefootnote{}\footnote{#1}%
  \addtocounter{footnote}{-1}%
  \endgroup
}

\renewcommand{\shortauthors}{Li et al.}

\begin{abstract}
  \blfootnote{$^{*}$Equal contribution. \quad $^{\dagger}$Equal advising.}%
  Instructional videos are the dominant medium for completing physical tasks, yet rarely match the user's visual context. %
  Motor simulation and cognitive load theories predict this mismatch matters, but how much, which attributes are responsible, and how users experience it remain unknown. %
  We conduct two studies (56 participants, two first-aid and two cooking tasks) using researcher-prepared recordings to control visual alignment. %
  Study~1 ($N{=}16$) compares fully aligned In-Context instructional videos (ICON) against business-as-usual Internet videos. ICON yields statistically significant gains: 11.1\% higher completion quality and 15.5\% faster completion. %
  Qualitative analysis decomposes the effect into four visual context attributes: Task Object Intrinsic, Task Object State, Environmental Context, and Observational Context. %
  Study~2 ($N{=}40$) ablates attributes one at a time. All four degrade objective performance yet users perceive no performance loss for any of them. %
  Visual context misalignment is \textit{substantial, decomposable, and invisible to the user for any single attribute}. %
\end{abstract}



\begin{teaserfigure}
  \centering
  \includegraphics[width=0.99\linewidth]{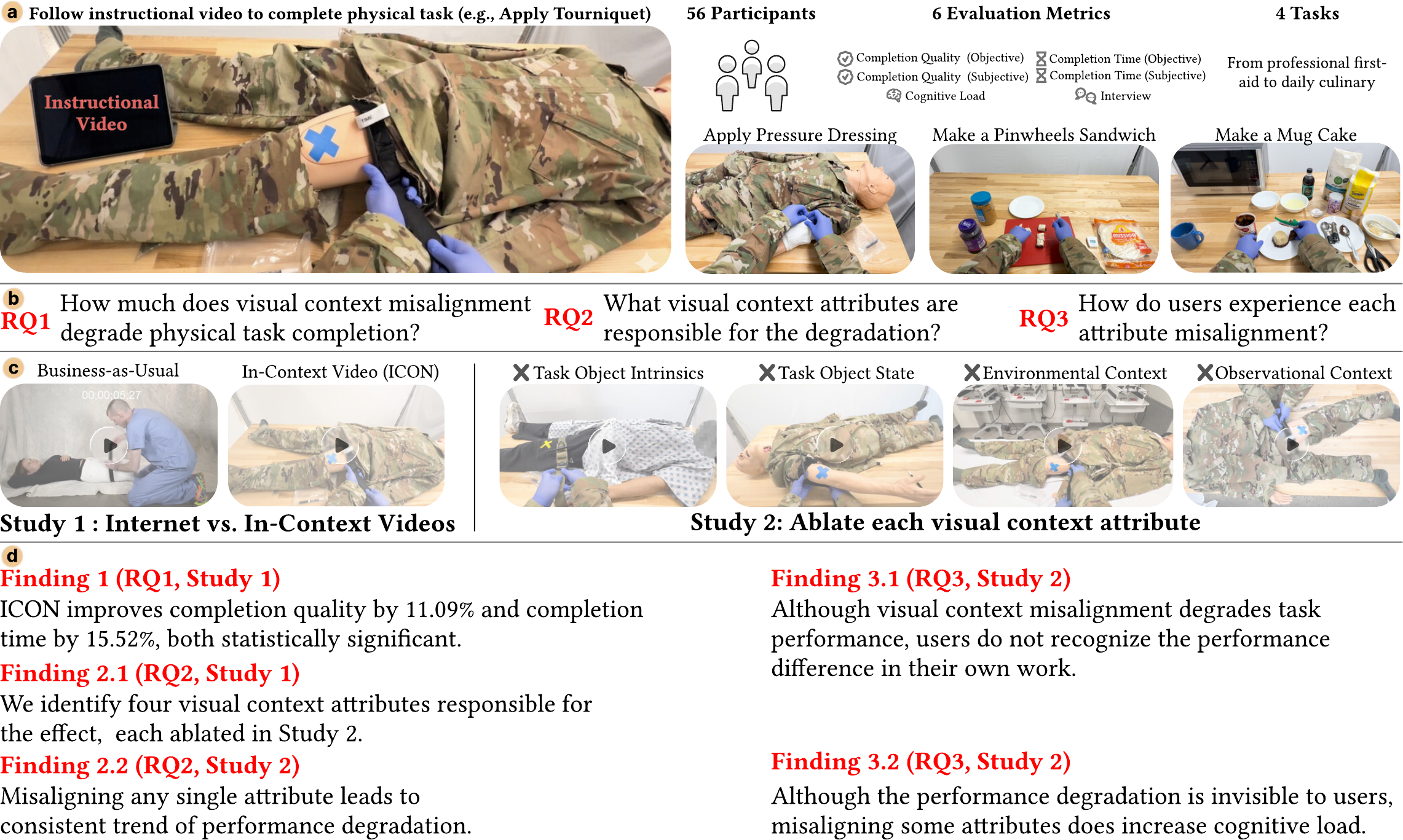}
  \caption{Instructional videos rarely match the viewer's own setup. We show the cost is substantial, decomposable, and invisible to the user. \textbf{(a) Problem.} Novices follow an instructional video whose visual context (the objects, their starting arrangement, the surroundings, and the viewpoint) differs from theirs. \textbf{(b) Research questions.} How much does the mismatch cost (RQ1), which attributes are responsible (RQ2), and how do users experience it (RQ3)? \textbf{(c) Studies.} Study~1 measures the whole cost (RQ1): Internet videos versus In-Context videos (ICON) recorded in the participant's setup. Study~2 isolates each attribute (RQ2, RQ3) by misaligning one at a time. \textbf{(d) Findings.} ICON raises completion quality and shortens completion time (RQ1: substantial). Misaligning any one of the four attributes degrades performance (RQ2: decomposable). Users do not notice the loss (RQ3: invisible).}
  \label{fig:teaser}
\end{teaserfigure}

\maketitle
\pagestyle{plain}
\thispagestyle{plain}
\section{Introduction}
Instructing unfamiliar physical tasks, from emergency first-aid to equipment assembly to cooking, touches millions of people: instructional videos account for a large share of online video consumption~\cite{smith2018pew}, and professional training programs increasingly rely on video demonstrations~\cite{morgado2024video}. %
Unlike software or digital tasks, physical tasks require coordinating body movements with variable real-world objects, tools, and environments, making effective instruction fundamentally harder to deliver~\cite{starcic2015transforming}. %
Given this complexity, video has become the dominant instructional medium for physical skills: it captures hand movements, object manipulations, and action sequences that text and static images cannot convey~\cite{mayer2020multimedia, sener2022assembly101}. %
As a result, understanding how the content of instructional videos affects physical task completion is scientifically important---it probes the link between visual perception and motor execution---and practically urgent, given how widely people rely on these videos for tasks with real consequences~\cite{youssef2022effect, wang2023holoassist}. %
Prior HCI research has improved how users navigate and follow instructional videos---through step segmentation~\cite{kim2013toolscape, chi2012mixt}, keyframe indexing~\cite{komlodi1998key}, and progress-aware playback~\cite{fraser2019replay}---but these systems offer limited insights on the case when visual context mismatches the user's situation. %
In practice, instructional videos on the Internet usually do not match the user's specific situation---different object models, different environments, different camera perspectives---and users must mentally translate between what they see on screen and what they face in reality. %
Cognitive science predicts that such mismatches should affect performance: motor simulation theory holds that visual similarity between observed and to-be-performed actions supports motor planning~\cite{jeannerod1994mental, jeannerod2001neural}, and cognitive load theory predicts that mentally translating across mismatched visual contexts imposes extraneous load that detracts from task execution~\cite{sweller1988cognitive, sweller2011cognitive}. %
However, these theoretical predictions leave three research questions open for real-world physical tasks:%

\begin{itemize}
\item \textbf{RQ1:} How much does visual context misalignment in instructional videos degrade physical task completion?
\item \textbf{RQ2:} What visual context attributes are responsible for this degradation?
\item \textbf{RQ3:} How do users experience each attribute's misalignment?
\end{itemize}

Answering RQ1 would establish the quantitative effect size that the field currently lacks. A growing body of work invests in AI-generated instructional content, from in-context image generation~\cite{lai2023lego, souvcek2024genhowto} to video synthesis~\cite{dai2023fine, chen2024videocrafter2, yang2024cogvideox}, yet whether visual context alignment produces a negligible or substantial effect remains unknown. An empirical estimate would ground these efforts. %
Answering RQ2 would identify which visual properties matter for task completion, providing a principled evaluation framework. Existing video quality metrics focus on fidelity and semantic coherence~\cite{huang2024vbench}, not on dimensions that predict instructional usefulness. A grounded attribute taxonomy could fill this gap. %
Answering RQ3 would test whether users can reliably detect misalignment's impact on their own performance. Metacognitive monitoring research suggests they may not~\cite{koriat2005illusions}. If so, the field's reliance on subjective ratings may systematically overestimate the quality of misaligned instructional content. %
Together, these questions address empirical gaps in a field where research interest and technical capability are advancing fast but the foundational knowledge is missing. %
To answer these questions, we use researcher-prepared instructional videos. Inspired by Wizard-of-Oz prototyping, in which a hand-made stand-in replaces a system that does not yet exist, we record the videos ourselves, before any session and under controlled conditions, to produce the level of visual alignment under investigation. %
Study~1 ($N{=}16$, within-subjects) is ecological. Each participant completes four physical tasks (two professional first-aid procedures and two everyday culinary tasks) twice: once with a popular Internet video for the task, the business-as-usual (BAU) condition, and once with our fully aligned recording, which we call an In-CONtext instructional video (ICON). This comparison keeps every difference a real video carries, so it measures the total gap between the status quo and full alignment (RQ1), the quantity anyone deciding whether to invest in alignment needs. It does not isolate visual context, and we do not read it as if it did. Its post-trial debriefings, in which participants who have just used both videos describe what they struggled with, yield the visual context attributes Study~2 tests (RQ2). %
Study~2 ($N{=}40$, between-subjects) isolates. Same enclosure, same demonstrator, same steps, one attribute misaligned per condition, and forty new participants each using one version. This measures each attribute's contribution on its own (RQ2) and lets us compare objective outcomes against participants' ratings of the video and of their own performance (RQ3). Production quality favors the baseline: ICON is a single unedited take narrated by a text-to-speech reading of the printed task sheet, while the BAU videos are edited and narrated by presenters. The one advantage ICON keeps beyond alignment, following the printed steps exactly, is shared by every Study~2 condition. %
Across both studies, 56 participants completed four tasks in over 86 hours of recorded trials. %
\textbf{Finding 1 (RQ1, Study~1):} Visual context misalignment carries a statistically significant performance cost. ICON outperforms BAU by 11.09\% in task completion quality and reduces completion time by 15.52\%, with a 21.1\% reduction in cognitive load, measured by the unweighted NASA-TLX score. In the tourniquet task, the average saving was about two minutes (329 s to 202 s), time that matters when bleeding is uncontrolled~\cite{kragh2009survival}. In the per-task comparisons, participants chose ICON over BAU 78.1\% of the time for completion quality and time and 79.7\% of the time for cognitive load. %
\textbf{Finding 2 (RQ2, Studies~1\&2):} The Study~1 debriefings name four visual context attributes responsible for the effect: Task Object Intrinsic (e.g., what the objects look like), Task Object State (e.g., how they are arranged at the start), Environmental Context (e.g., the surroundings), and Observational Context (e.g., the camera's viewpoint). In Study~2, misaligning any one of them alone from an otherwise matched video lowered completion quality, by 4.52\% to 9.83\%, with Task Object State and Environmental Context reaching significance, and lengthened completion time in every case, though those differences fell short of significance. %
\textbf{Finding 3 (RQ3, Study~2):} This cost is invisible to the people paying it. Participants rated every misaligned video as easy to follow and as helpful as ICON; the group with the largest quality drop gave the most ``strongly agree'' answers to ``the video improves my task performance quality.'' Neither weighted NASA-TLX scores nor the video and performance ratings differed significantly between ICON and any ablation condition. When we pointed out the mismatch afterwards, participants said they had not noticed it (\user{Oh, I didn't even notice that}, P7) or had noticed and assumed it did not matter. %
Viewers can tell when everything differs at once: Study~1 participants, who used both videos, chose ICON four times out of five. A single misaligned attribute, though, leaves the video feeling fluent, and people read fluency as adequate performance~\cite{oppenheimer2008metacognition}. A viewer's rating therefore cannot tell whether an instructional video fits its viewer. Only measured performance can.

\section{Related Work}

\subsection{Instructional Videos for Physical Task Guidance}
People perform unfamiliar physical tasks every day, and instructional videos are the dominant way they learn: how-to videos account for a large share of online video consumption~\cite{smith2018pew}, and professional training in surgery~\cite{youssef2022effect}, manufacturing~\cite{sener2022assembly101}, and emergency response increasingly relies on video demonstrations~\cite{morgado2024video}. The scale and real-world consequences of this use make research on instructional video for physical tasks both scientifically important and practically urgent. %
HCI research has made these videos easier to use---through step segmentation~\cite{kim2013toolscape, chi2012mixt}, keyframe indexing~\cite{komlodi1998key}, search~\cite{kim2023surch}, comprehension aids~\cite{yang2023videomap, aftab2020remo, yang2024aqua, zhong2021helpviz}, and multi-video aggregation~\cite{yang2025videomix}---but this work focuses on navigation and understanding of the depicted content~\cite{yang2023videomap}, not on what happens when that content does not match the user’s environment. Context-adaptive systems exist for software tutorials~\cite{fraser2019replay, wang2014evertutor, perraud2024tutorial} and AR-based physical guidance~\cite{huang2021adaptutar, liu2023instrumentar}, but they address step-level matching (which step is the user on?) or overlay registration, not whether the depicted objects, environment, and viewpoint actually match what the user faces. What happens when this visual context does not match remains unstudied for physical tasks. %

\subsection{Cognitive Science of Visual Context Mismatch}%
Cognitive science offers directional predictions that visual context mismatch should hurt performance. Motor simulation theory holds that observing an action triggers internal motor planning whose fidelity depends on visual similarity between the demonstration and the observer’s own situation~\cite{jeannerod1994mental, jeannerod2001neural}: when objects, viewpoint, and body movements match, motor planning is direct; when they differ, the observer must mentally transform the demonstration to map it onto their own context. Cognitive load theory provides a complementary prediction---this mental translation constitutes extraneous load that consumes limited working memory at the expense of task execution~\cite{sweller1988cognitive, sweller2011cognitive, chandler1991cognitive}. Prior work has confirmed that video design choices affect cognitive load across dimensions such as explanation clarity, information redundancy, and goal alignment~\cite{Fan2024Video, Wang2020Converging, Costley2020The, Yoon2022Effects, Juliano2021Increased}. However, these predictions and findings come from controlled laboratory paradigms studying simple, single-action tasks. The closest test in instructional video is Fiorella et al.'s finding that a circuit-assembly task is learned better from a first-person recording than from a third-person one~\cite{fiorella2017perspective}, a single attribute in isolation. No prior work has decomposed what ``visual context’’ means for multi-step physical task completion, tested more than one of its attributes, or investigated how users themselves experience the mismatch. %
\subsection{Metacognition and Change Blindness}
Cognitive science also predicts that even when visual context misalignment degrades performance, users may not notice. People use processing fluency---how easily material is processed---as a heuristic for their own performance quality~\cite{oppenheimer2008metacognition}. Koriat and Bjork~\cite{koriat2005illusions} demonstrated this as an ``illusion of competence’’: conditions that feel fluent produce overconfidence despite objectively worse outcomes. Change blindness compounds the effect: observers routinely miss substantial visual changes outside their attentional focus~\cite{simons1999gorillas}, and when following an instructional video, users attend to task progress (e.g., what to do next, which tool to grab) not to whether the depicted objects or environment match their own. Together, these mechanisms predict that near-aligned videos will feel fluent enough that users lack any subjective signal of degradation. Our Finding~3, in which participants fail to perceive the effect of single-attribute misalignment despite measurable performance drops, provides empirical evidence for exactly this prediction in the practically important domain of physical task instruction. %

\begin{figure*}[t]
    \centering
    \includegraphics[width=\textwidth]{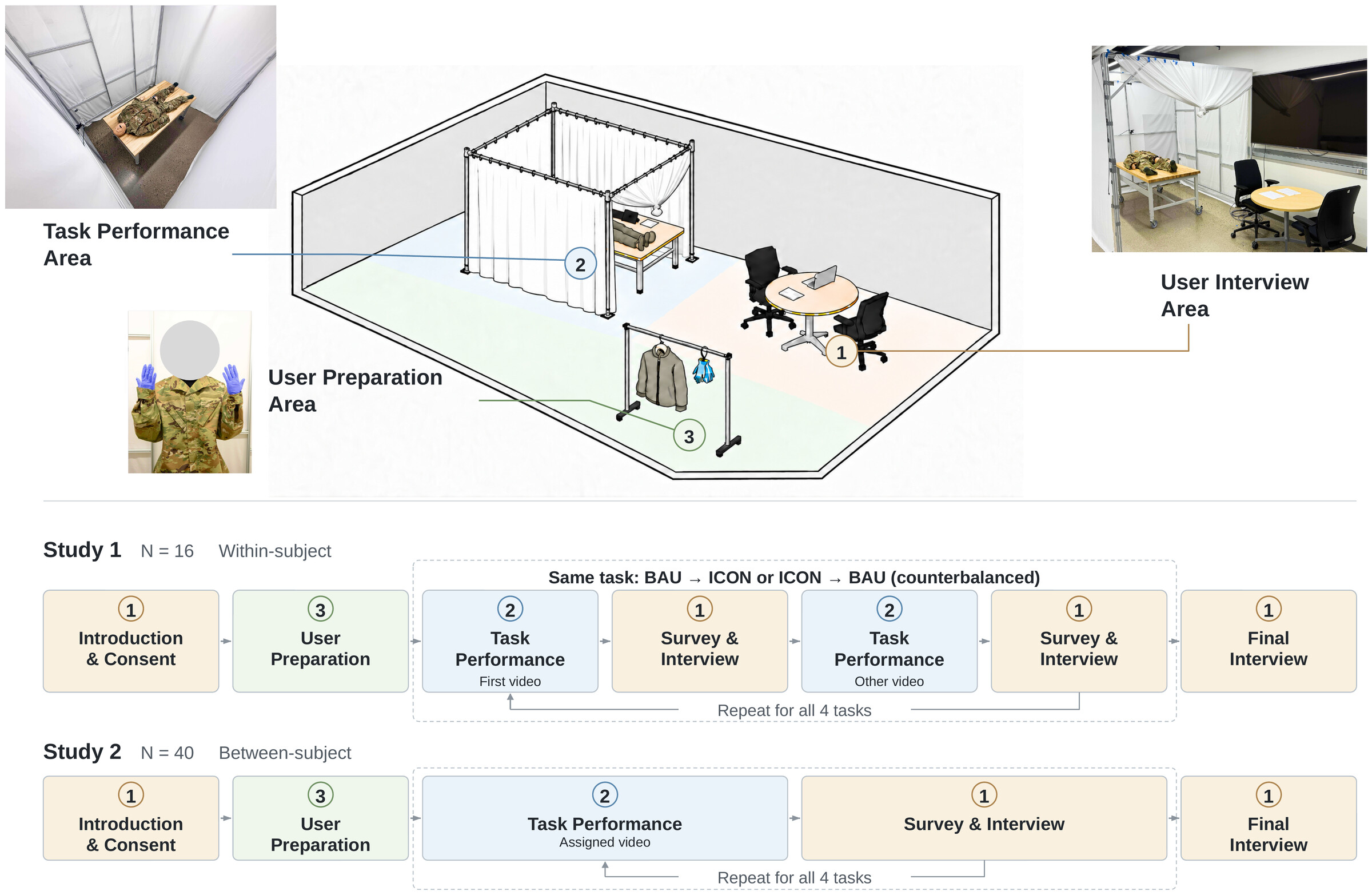}
    \caption{Study setting and procedures. \textbf{Top}: The room illustration identifies the user interview area (1), task performance area (2), and user preparation area (3). Pictures show the enclosed task area (upper left), the room overview (upper right), and the standardized clothing and gloves worn by participants and research personnel (lower left). The enclosure controls visual context and limits external visual distractions, while standardized clothing and gloves provide a consistent first-person visual experience. The front curtain is shown open for illustration and closed during trials. \textbf{Bottom}: Study 1 uses a within-subject design, with each task performed using both BAU and ICON in counterbalanced order. Study 2 uses a between-subject design, with one video condition assigned to each participant. Each trial is followed by a survey and interview. Dashed outlines mark the steps repeated for all four tasks, and both studies end with a final interview. Numbered circles indicate room areas.}
    \label{fig:study-procedure}
    \Description{A room schematic above two study flows. A round table supports interviews, a curtained workspace supports task performance, and a clothes rack marks the preparation area. Study 1 repeats each task with both video conditions; Study 2 uses one assigned condition per participant. Both studies include preparation, trial-level surveys and interviews, and a final interview.}
\end{figure*}

\section{Method Overview}
In this paper, we compare how different levels of visual context alignment in instructional videos impact successful assistance in carrying out unfamiliar physical tasks among novice users.
In both user studies, we recruit participants to perform a set of physical tasks that each have associated instructional videos from different visual context settings.
This section outlines the shared methodologies and settings for both studies, including the physical infrastructure, procedural consistency, task selection, participant criteria, and evaluation metrics.

To ensure precise control over the visual context experienced by participants, we construct a dedicated ``enclosure'' that serves as the study area (Fig.~\ref{fig:study-procedure}). This enclosed setup separates the task performance area from the communication area, and is open on top to avoid confinement. This setup standardizes all visual elements involved: participants wear uniform clothing and gloves, and receive instructional video on a shared tablet. This consistency in setup is crucial for attributing differences in task outcomes solely to the type of instructional video used, whether BAU, ICON, or ablated ICONs. %

\subsection{User Study Procedure}
Each participant begins by receiving a briefing and providing informed consent. Participants are then outfitted with standardized clothing and gloves, and given uniform instructions to ensure a shared understanding of the task requirements. Their goal is to complete the task described on the paper as accurately and efficiently as possible, assisted by the instructional videos provided on a tablet with a YouTube player. Participants have the flexibility to choose how they engage with the materials, whether by watching the video first or simultaneously with the task, and can control playback as needed.
After confirming their understanding, participants perform a trial in the task area. In each trial, one participant completes a task once with one type of video assistance, verbally marking the start and finish. After each trial, participants complete a brief survey and participate in a short debrief to provide qualitative feedback. A post-study interview is conducted to gather additional insights into their overall experience. Our two studies follow two different schedules of trials, both counterbalanced for order, achieving different research goals (overview in Fig.~\ref{fig:study-procedure}, details in Sec.~\ref{sec:us1} and Sec.~\ref{sec:us2}).

\subsection{Selected Physical Tasks}

We select four tasks spanning two domains: the professional medical domain (``Apply Tourniquet'' and ``Apply Pressure Dressing'') and the everyday cooking domain (``Make a Pinwheel Sandwich'' and ``Make a Mug Cake''), illustrated in Fig.~\ref{fig:bau_icon}. The detailed task description is provided in the Supplementary Material (Section~S1, Fig.~S1).
The tasks involve different types of objects and actions, but each requires sufficiently intricate physical manipulation to necessitate familiarization and assistance via spatiotemporal video instruction. ``Apply Tourniquet'' and ``Apply Pressure Dressing'' involve a small number of objects that require special training and gross motor skills. For example, ``Apply Tourniquet'' requires users to lift the leg and wrap the tourniquet in a specific way.
On the other hand, ``Make a Pinwheel Sandwich'' and ``Make a Mug Cake'' involve more objects, but they are commonly seen in daily life, and they require more fine-grained finesse to manipulate. For example, ``Make a Pinwheel Sandwich'' requires users to slice a tortilla roll with a floss into pieces.
All these tasks are performed by every participant in both user studies.

\subsection{Participant Selection}
We recruited participants from our university community through email and messaging-channel announcements and referrals; no one took part in both studies. We selected participants who had never performed any of the chosen tasks before. This aligns with our focus on video assistance for unfamiliar physical tasks. We used a prescreening background form to gather information about participants' familiarity with the tasks, rated on a five-level scale from never having heard of the task to having performed it multiple times, together with their habits of using how-to videos. All participants confirmed that they had ``never performed this task before'' for each of the four tasks under investigation. Selected participants were asked not to look into the tasks before their session, and their familiarity was reconfirmed upon arrival to ensure no additional knowledge was acquired between screening and participation.
This selection process ensures that our findings reflect the impact of video assistance on novice users. Both studies were approved by our institution's Institutional Review Board (IRB).

\subsection{
Evaluation Dimensions and Measures}
We assess how alignment of visual context in instructional videos of unfamiliar physical task affects task completion outcomes. We define how we evaluate the effects as follows.

\textbf{Task Completion Quality.}
\underline{Completion Quality Score}: We employ three experts, each with experience successfully completing the tasks over ten times, to grade video recordings using a detailed rubric. Prior to independent grading, the three experts jointly developed the rubric and conducted a calibration session on a randomly selected subset of two participants' recordings, achieving Fleiss' $\kappa = 1.0$ \cite{fleiss1971measuring}. All remaining recordings were then graded independently. Each step is graded on a scale from 0 to 1 (0: not done, 1: correctly done, with partial scores as per the rubric). The final grade is the average of all step grades.
\underline{Perceived Completion Quality}: Participants rate their agreement on a 7-level scale regarding whether the video improved task quality after each trial---Absolute Score. Participants also select which video they believe most enhanced their completion quality---Comparative Score, in within-subjects experiments (Study 1).

\textbf{Task Completion Time.}
\underline{Completion Time}: We record task completion time from the participant's verbal start to finish. Step completion time is annotated by the same experts, following the rubric.
\underline{Perceived Completion Speed}: Participants rate their agreement on a 7-point scale regarding whether the video helped save time---Absolute Score, and choose the most effective video in saving time---Comparative Score in within-subjects experiments (Study 1).

\textbf{User Cognitive Load.} We use the NASA-TLX \cite{hart1988development} instrument with a 10-point scale after each trial. In Study 2, participants complete 15 pairwise comparisons among the six dimensions to determine their weights. Study~1 reports the unweighted overall score together with the six subscales; Study~2 reports the weighted overall score. All reported NASA-TLX scores, including the displayed Performance subscale, are oriented so that higher scores indicate greater workload.

\textbf{Qualitative Interview Feedback.} To understand the reasons behind the quantitative data, we conduct interviews after each trial and at the end of all trials. Questions are designed to explore participants' general experiences, preferences for different video types, and their opinions on misaligned visual context. Following this principle, two studies have different questions designed (details in Sec.~\ref{sec:us1} and Sec.~\ref{sec:us2}). The complete question lists for both studies are in Supplementary Section~S3.

\section{Study 1: Comparing In-context with Business-as-usual Instructional Videos}
\label{sec:us1}

In Study~1, participants were instructed to strictly follow the paper instructions and use the assigned video as a reference. We compared two video conditions: Internet-sourced business-as-usual (BAU) videos (Fig.~\ref{fig:bau_icon}-left) and researcher-created In-Context Instructional Videos (ICON) (Fig.~\ref{fig:bau_icon}-right).

This study has two primary goals. First, we validate the general effectiveness of ICON over BAU and measure the total performance gap between the status quo and full alignment (\textbf{RQ1}); this study is ecological and does not isolate visual context, which Study~2 does. Second, we identify the key visual context attributes that drive ICON's effectiveness through qualitative analysis of participant interviews (\textbf{RQ2}).

\begin{table*}[t]
\captionsetup{position=bottom}
\centering
\begingroup
\small
\setlength{\tabcolsep}{6pt}
\renewcommand{\arraystretch}{1.15}
\begin{tabularx}{\textwidth}{@{}>{\raggedright\arraybackslash}p{.18\textwidth}*{2}{>{\raggedright\arraybackslash}X}@{}}
\toprule
\textbf{Characteristic} & \textbf{BAU} & \textbf{ICON} \\
\midrule
Source & Public videos selected from popular search results. & Videos of researchers performing the task. \\
\addlinespace[3pt]
Materials and setup & Tools, ingredients, and settings differ from those used in the study. & Same tools, materials, initial arrangement, and workspace as participants. \\
\addlinespace[3pt]
Task steps & Original procedures differ from the paper instructions. First-aid videos show a brief full demonstration followed by detailed teaching. & Follow the paper instructions exactly, in the specified order. \\
\addlinespace[3pt]
Narration & Conversational speech describing what to do and why. & TTS reading of the paper instructions only; no added explanations. \\
\addlinespace[3pt]
Pacing and editing & Normal-speed actions; any original cuts retained. Full videos used. & Normal-speed actions; continuous, uncut recordings. \\
\addlinespace[3pt]
Camera framing & Third-person view facing the demonstrator; framing may vary between close-ups and the whole workspace. & Fixed first-person view; no camera movement or framing changes. \\
\addlinespace[3pt]
Duration (s) & 139 / 99 / 435 / 160 & 84 / 85 / 381 / 614 \\
\bottomrule
\end{tabularx}
\par\vspace{4pt}
\begin{minipage}{\textwidth}\footnotesize
Duration order: Tourniquet / Pressure Dressing / Pinwheels / Mug Cake.
\end{minipage}
\endgroup
\caption{Characteristics of the BAU and ICON videos used in Study 1. BAU videos are unmodified Internet instructional videos; ICON videos are researcher-created ideal cases matched to the paper instructions and participants' setup. The table shows observed differences between BAU and ICON.}
\label{tab:us1-video-comparison}
\end{table*}

\subsection{Business-as-usual and In-context Instructional Videos}

BAU videos served as real-world references that users might encounter when searching for help with unfamiliar physical tasks. For each task, we searched YouTube using task-related keywords and selected a popular instructional video. We used these videos in full, without editing their content or introducing mismatches. The selected videos are listed below, with view counts recorded at submission:
\href{https://youtu.be/tOm8lJLRWF0}{Tourniquet}%
\footnote{Tourniquet (300,377 views):
\url{https://youtu.be/tOm8lJLRWF0}},
\href{https://youtu.be/LdXZ2gwIqbA}{Pressure Dressing}%
\footnote{Pressure Dressing (81,904 views):
\url{https://youtu.be/LdXZ2gwIqbA}},
\href{https://youtu.be/Nj_Y-SovfwM}{Pinwheels}%
\footnote{Pinwheels (1,007,900 views):
\url{https://youtu.be/Nj_Y-SovfwM}},
and \href{https://youtu.be/DWKp7zLKWwg}{Mug Cake}%
\footnote{Mug Cake (2,502,667 views):
\url{https://youtu.be/DWKp7zLKWwg}}.

ICON videos were created to match to the paper instructions and participants' physical setup. A demonstrator with extensive experience in the tasks performed the recordings in our experimental enclosure (Fig.~\ref{fig:study-procedure}). The demonstrations followed the paper instructions exactly, using the same tools, materials, initial arrangement, and workspace as participants. All ICON videos used a fixed first-person view, with no camera movement or framing changes. Narration consisted only of a text-to-speech (TTS) reading of the paper instructions, with no additional explanation.

\begin{figure}[t]
  \centering
  \includegraphics[width=\linewidth,height=0.9\textheight,keepaspectratio]{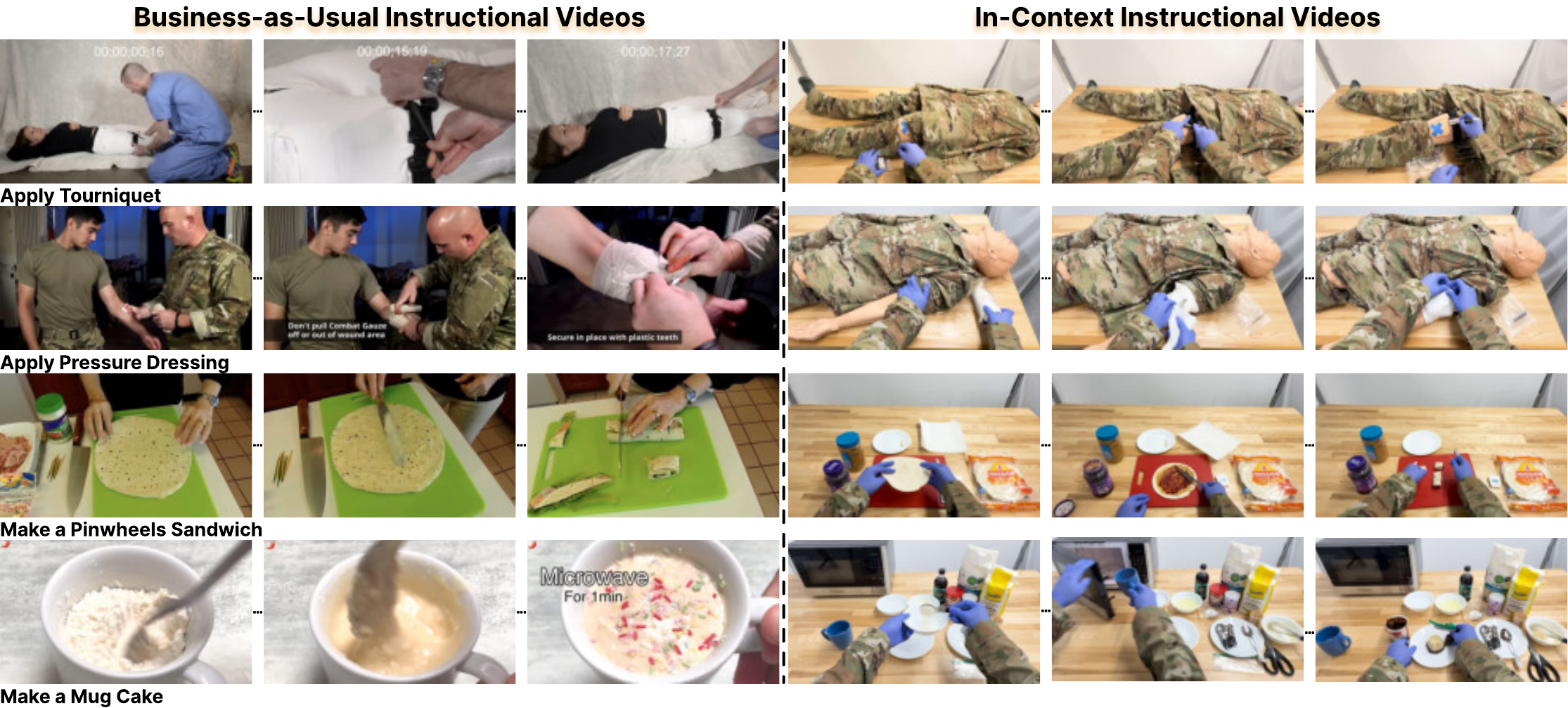}
  \caption{The two video types compared in Study 1: Business-as-Usual (BAU) and In-Context (ICON) instructional videos. Key frames show the beginning (left), middle (middle), and ending (right) stage of each task under each video type.
  }
  \label{fig:bau_icon}
    \vspace{-15pt}
\end{figure}

Video selection and preparation were not intended to favor either condition. Figure~\ref{fig:bau_icon} shows key frames from both conditions, and Table~\ref{tab:us1-video-comparison} summarizes their characteristics. These differences reflect the selected Internet videos and the researcher-created ideal cases; individual video features were not separately manipulated in Study 1.

\subsection{Experiment Design and Study Procedure}

Each participant experiences both video conditions (BAU and ICON) for each task, following a within-subject design where each task is performed twice. This design allows us to gather both quantitative metrics for effectiveness validation, for \textbf{RQ1}. Also, participants provide comparative insights through interviews about key visual context attributes that drive ICON's effectiveness, for \textbf{RQ2}.

To counterbalance task and video-condition order, we crossed four cyclic task orders with two video orders (BAU followed by ICON, or ICON followed by BAU). This yielded eight trial sequences, with two participants assigned to each sequence (Table~\ref{tab:us1-trial-order}).

\begin{table*}[t]
\centering
\begingroup
\fontsize{7}{8}\selectfont\sffamily
\setlength{\tabcolsep}{3pt}
\renewcommand{\arraystretch}{1.0}

\renewcommand{\ICONRevBodyWidthTQ}{26pt}
\renewcommand{\ICONRevBodyWidthPD}{20pt}
\renewcommand{\ICONRevBodyWidthPW}{20pt}
\renewcommand{\ICONRevBodyWidthMC}{16pt}

\renewcommand{\ICONRevTrialCell}[2]{%
  \cellcolor{ICONRev#1#2}%
  \begin{minipage}[c][31pt][c]{\linewidth}
    \centering
    \parbox[c][18pt][c]{\linewidth}{%
      \centering
      \csname ICONRev#1\endcsname{%
        \csname ICONRevBodyWidth#1\endcsname}%
    }\par
    \vspace{1pt}%
    {\sffamily\bfseries\fontsize{7}{8}\selectfont #2}
  \end{minipage}%
}

\arrayrulecolor{ICONRevGrid}
\begin{tabularx}{\textwidth}{@{}>{\centering\arraybackslash}m{.12\textwidth}*{8}{>{\centering\arraybackslash}X}@{}}
\toprule
\rowcolor{ICONRevHeader}
\textbf{Participant} & \textbf{Trial 1} & \textbf{Trial 2} & \textbf{Trial 3} & \textbf{Trial 4} & \textbf{Trial 5} & \textbf{Trial 6} & \textbf{Trial 7} & \textbf{Trial 8} \\
\midrule
\textbf{P1} & \ICONRevTrialCell{TQ}{BAU} & \ICONRevTrialCell{TQ}{ICON} & \ICONRevTrialCell{PD}{BAU} & \ICONRevTrialCell{PD}{ICON} & \ICONRevTrialCell{PW}{BAU} & \ICONRevTrialCell{PW}{ICON} & \ICONRevTrialCell{MC}{BAU} & \ICONRevTrialCell{MC}{ICON} \\
\textbf{P2} & \ICONRevTrialCell{TQ}{ICON} & \ICONRevTrialCell{TQ}{BAU} & \ICONRevTrialCell{PD}{ICON} & \ICONRevTrialCell{PD}{BAU} & \ICONRevTrialCell{PW}{ICON} & \ICONRevTrialCell{PW}{BAU} & \ICONRevTrialCell{MC}{ICON} & \ICONRevTrialCell{MC}{BAU} \\
\hline
\textbf{P3} & \ICONRevTrialCell{PD}{BAU} & \ICONRevTrialCell{PD}{ICON} & \ICONRevTrialCell{PW}{BAU} & \ICONRevTrialCell{PW}{ICON} & \ICONRevTrialCell{MC}{BAU} & \ICONRevTrialCell{MC}{ICON} & \ICONRevTrialCell{TQ}{BAU} & \ICONRevTrialCell{TQ}{ICON} \\
\textbf{P4} & \ICONRevTrialCell{PD}{ICON} & \ICONRevTrialCell{PD}{BAU} & \ICONRevTrialCell{PW}{ICON} & \ICONRevTrialCell{PW}{BAU} & \ICONRevTrialCell{MC}{ICON} & \ICONRevTrialCell{MC}{BAU} & \ICONRevTrialCell{TQ}{ICON} & \ICONRevTrialCell{TQ}{BAU} \\
\hline
\textbf{P5} & \ICONRevTrialCell{PW}{BAU} & \ICONRevTrialCell{PW}{ICON} & \ICONRevTrialCell{MC}{BAU} & \ICONRevTrialCell{MC}{ICON} & \ICONRevTrialCell{TQ}{BAU} & \ICONRevTrialCell{TQ}{ICON} & \ICONRevTrialCell{PD}{BAU} & \ICONRevTrialCell{PD}{ICON} \\
\textbf{P6} & \ICONRevTrialCell{PW}{ICON} & \ICONRevTrialCell{PW}{BAU} & \ICONRevTrialCell{MC}{ICON} & \ICONRevTrialCell{MC}{BAU} & \ICONRevTrialCell{TQ}{ICON} & \ICONRevTrialCell{TQ}{BAU} & \ICONRevTrialCell{PD}{ICON} & \ICONRevTrialCell{PD}{BAU} \\
\hline
\textbf{P7} & \ICONRevTrialCell{MC}{BAU} & \ICONRevTrialCell{MC}{ICON} & \ICONRevTrialCell{TQ}{BAU} & \ICONRevTrialCell{TQ}{ICON} & \ICONRevTrialCell{PD}{BAU} & \ICONRevTrialCell{PD}{ICON} & \ICONRevTrialCell{PW}{BAU} & \ICONRevTrialCell{PW}{ICON} \\
\textbf{P8} & \ICONRevTrialCell{MC}{ICON} & \ICONRevTrialCell{MC}{BAU} & \ICONRevTrialCell{TQ}{ICON} & \ICONRevTrialCell{TQ}{BAU} & \ICONRevTrialCell{PD}{ICON} & \ICONRevTrialCell{PD}{BAU} & \ICONRevTrialCell{PW}{ICON} & \ICONRevTrialCell{PW}{BAU} \\
\bottomrule
\end{tabularx}
\par\vspace{5pt}
{\footnotesize
\setlength{\tabcolsep}{4pt}
\setlength{\fboxsep}{3pt}

\begin{tabularx}{\textwidth}
{@{}*{4}{>{\centering\arraybackslash}X}@{}}

\raisebox{-.2\height}{\ICONRevTQ{22pt}}~Tourniquet &
\raisebox{-.2\height}{\ICONRevPD{17pt}}~Pressure Dressing &
\raisebox{-.2\height}{\ICONRevPW{17pt}}~Pinwheels &
\raisebox{-.2\height}{\ICONRevMC{14pt}}~Mug Cake
\\[4pt]

\colorbox{ICONRevTQBAU}{\strut BAU}\quad
\colorbox{ICONRevTQICON}{\strut ICON}
&
\colorbox{ICONRevPDBAU}{\strut BAU}\quad
\colorbox{ICONRevPDICON}{\strut ICON}
&
\colorbox{ICONRevPWBAU}{\strut BAU}\quad
\colorbox{ICONRevPWICON}{\strut ICON}
&
\colorbox{ICONRevMCBAU}{\strut BAU}\quad
\colorbox{ICONRevMCICON}{\strut ICON}
\\

\end{tabularx}
\par}
\arrayrulecolor{black}
\endgroup
\caption{Study 1 trial orders for P1--P8. Each row shows eight consecutive trials: each of four tasks is performed twice, once with BAU and once with ICON. Four cyclic task orders are crossed with two video orders. Colors identify tasks; lighter and darker shades indicate BAU and ICON, respectively. P9--P16 repeat the sequences of P1--P8.}
\label{tab:us1-trial-order}
\end{table*}

Given the two conditions (BAU and ICON) and four tasks, each participant completes 8 trials, resulting in a 2 to 2.5-hour session, with 1.5 to 2 hours dedicated to task performance and the rest for post-trial surveys and interviews. We recruited 16 participants, each compensated \$60, yielding 35 hours of task performance data for analysis.

In this user study, after each trial, participants complete a NASA-TLX form and a Likert scale assessment. After every two trials---where participants experienced two different methods on the same task---a debriefing session was conducted to gather their opinions on comparing the two types of videos. Following the completion of all trials, participants were interviewed to gather their insights on which visual context attributes they deemed important in video assistance (question list in Supplementary Section~S3).

\subsection{Finding 1: Evaluating ICON Effectiveness over BAU}
Addressing \textbf{RQ1}, we built a linear mixed-effects model with \textit{Score} as the dependent variable, \textit{Method} (ICON vs. BAU), \textit{Task}, and their interaction (\textit{Method} $\times$ \textit{Task}) as fixed effects, and a random intercept for each participant to account for individual differences. This model allows us to evaluate the overall impact of the instructional method on different dimensions, as well as whether this effect varies across different tasks. The formula is shown below, and Table~\ref{tab:us1_lmm_omnibus} reports the omnibus tests from this model for each metric.

\begin{equation}
\textit{Score} = \textit{Method} * \textit{Task} + (1|\textit{Participant})
\end{equation}

\textbf{Overall, we found that In-Context Instructional Videos (ICON) improved task completion quality, decreased completion time, and reduced cognitive load when compared to business-as-usual (BAU) videos, with no significant Method $\times$ Task interaction on any of the three measures.} Participants consistently favored ICON videos, highlighting their superior alignment with the visual context and the reduced necessity for mental mapping.

\begin{figure*}
  \centering
  \includegraphics[width=0.8\textwidth]{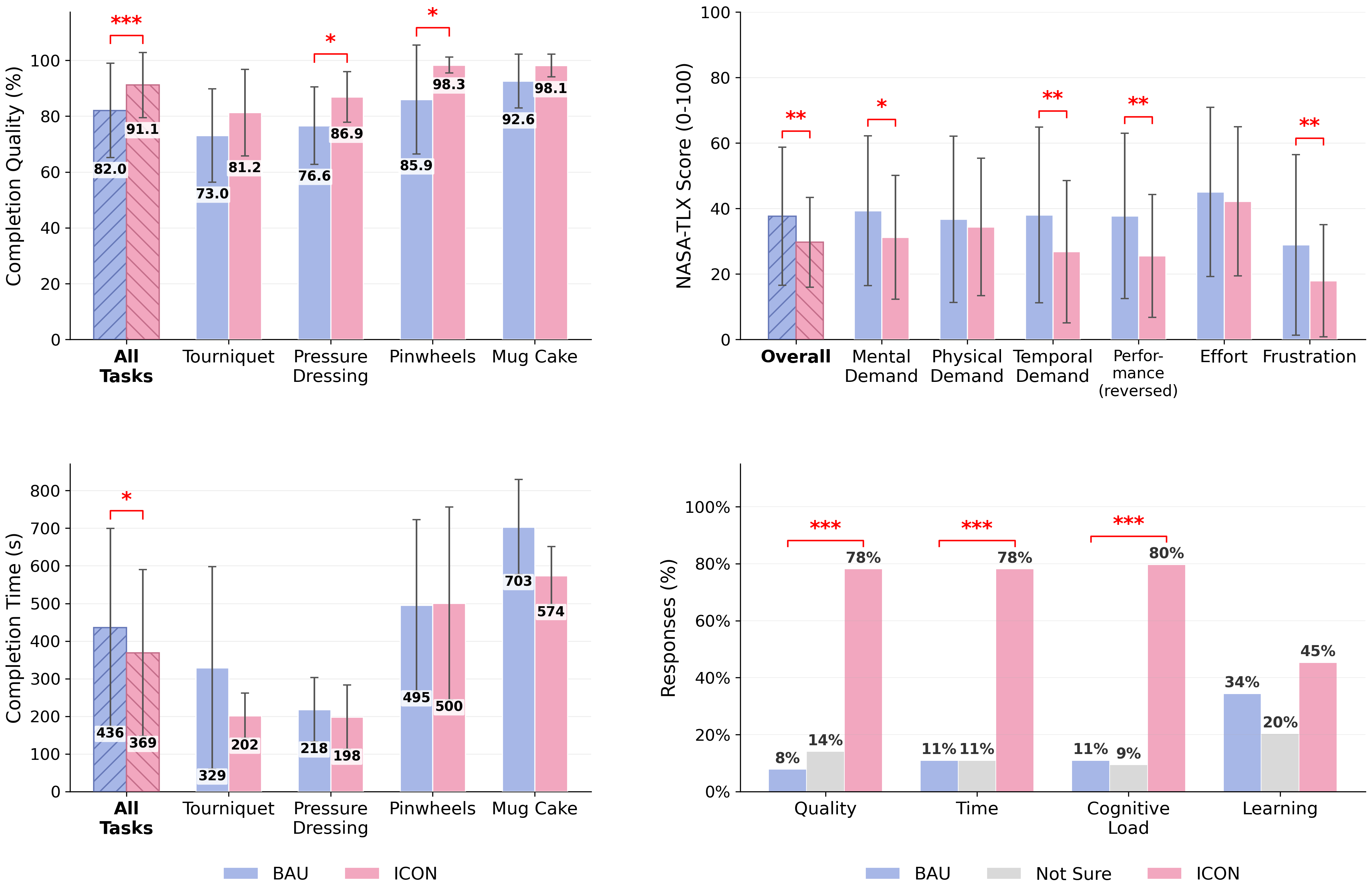}
  \caption{Comparison of task performance, subjective workload, and video preferences between BAU and ICON in Study 1. \textbf{Top left:} Completion quality (\%). \textbf{Bottom left:} Completion time (s). Both panels show results across all tasks and for each task. \textbf{Top right:} Unweighted overall NASA-TLX and individual subscale scores (0--100), with Performance reverse-coded so that higher scores indicate greater workload. \textbf{Bottom right:} Participants' preferences for which condition improved quality, saved time, reduced cognitive load, and enhanced learning. Bars in the quality, time, and NASA-TLX panels show means; error bars show trial-level standard deviations. Statistical significance is denoted by * ($p < 0.05$), ** ($p < 0.01$), and *** ($p < 0.001$). Quality, time, and NASA-TLX comparisons use the interaction LMMs with a participant random intercept. Holm adjustment is applied across the four task-specific comparisons separately for quality and time, and across the six NASA-TLX subscales; each overall contrast is tested separately without adjustment. Preference comparisons test ICON versus BAU using intercept-only multinomial logistic models, retaining all three response categories, without multiplicity adjustment.}
  \label{fig:us1_obj}
\end{figure*}

\providecommand{\sigp}[2]{\textbf{#1\textsuperscript{#2}}}

\begin{table*}[t]
  \centering
  \small
  \begin{tabular}{lrrrrrr}
    \toprule
    & \multicolumn{3}{c}{Main Effect of Method}
    & \multicolumn{3}{c}{Method $\times$ Task Interaction} \\
    \cmidrule(lr){2-4}\cmidrule(lr){5-7}
    Metric & $\chi^2$ & df & $p$ & $\chi^2$ & df & $p$ \\
    \midrule
    Completion Quality (\%) $\uparrow$
      & 19.32 & 1 & $\mathbf{<0.001}^{***}$ & 1.48 & 3 & 0.688 \\
    Completion Time (s) $\downarrow$
      & 5.70 & 1 & \textbf{0.017}\textsuperscript{*}
      & 4.64 & 3 & 0.200 \\
    Unweighted NASA-TLX (0--100) $\downarrow$
      & 9.62 & 1 & $\mathbf{0.002}^{**}$ & 2.27 & 3 & 0.518 \\
    \bottomrule
  \end{tabular}
  \caption{Linear mixed-model analysis results for Study~1. One model was fitted per metric using restricted maximum likelihood: $\mathrm{Metric}\sim\mathrm{Method}\times\mathrm{Task}+(1\mid\mathrm{Participant})$, with BAU as the reference method. Both sets of columns report Wald $\chi^2$ tests from this interaction model. The main effect of Method tests the average marginal contrast of ICON versus BAU, with equal weight across the four tasks (df = 1); it is the squared Wald $z$ for that contrast. The Method $\times$ Task test jointly tests the three interaction coefficients (df = 3). NASA-TLX Performance was reverse-scored before normalization to 0--100 and averaging across the six subscales, so that higher overall scores indicate greater workload. Significant $p$-values are bolded; $^{*}p<0.05$, $^{**}p<0.01$, and $^{***}p<0.001$.}
  \label{tab:us1_lmm_omnibus}
\end{table*}

\begin{figure*}
  \centering
  \includegraphics[width=\textwidth]{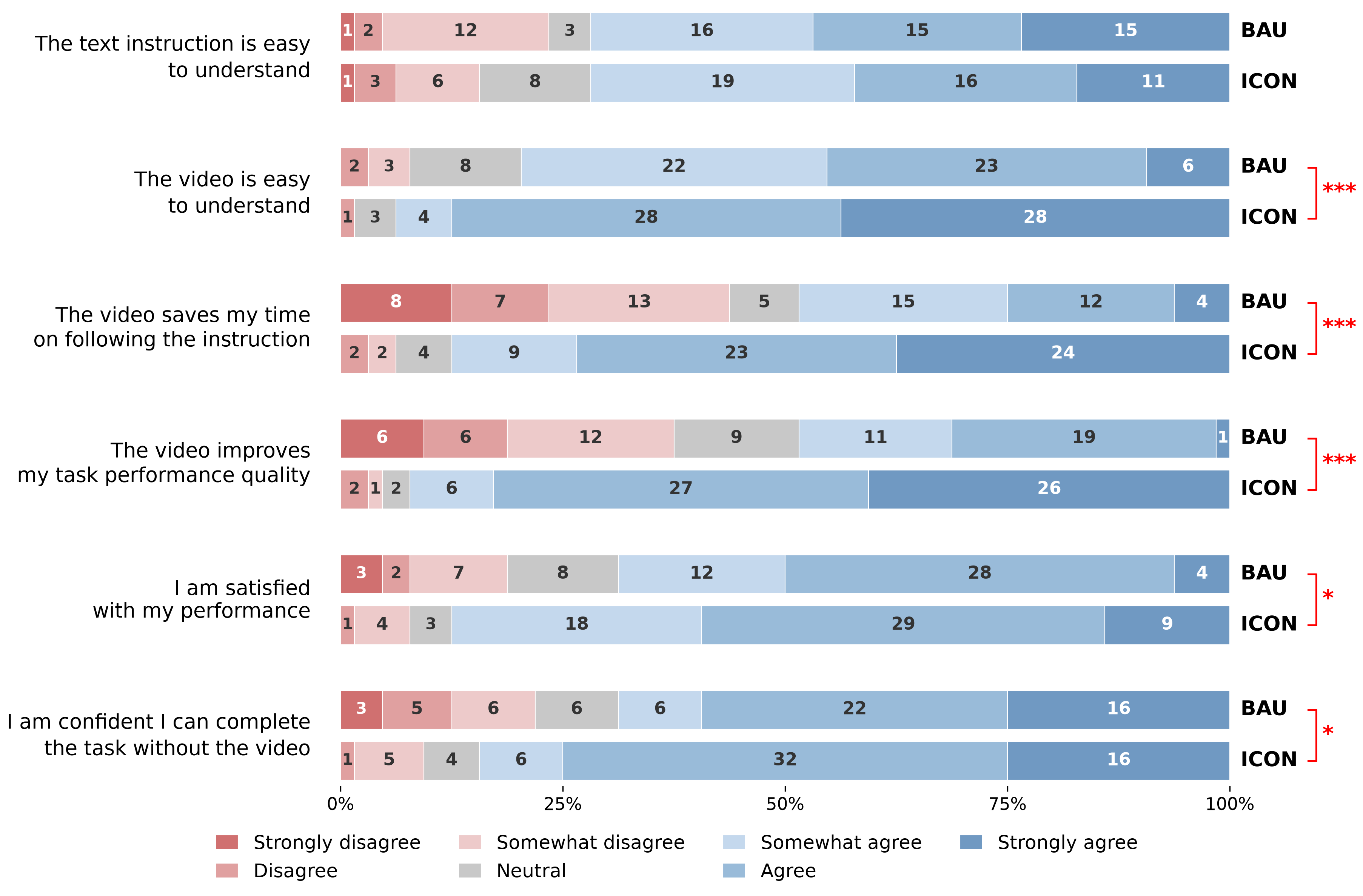}
  \caption{Study 1 Likert-scale responses comparing BAU and ICON on six subjective measures: text instruction clarity, video understandability, perceived time savings, perceived task performance quality improvement, satisfaction, and confidence in completing tasks without video. Bar widths indicate response proportions, and numbers within segments indicate response counts. Questions appear on the left, with video conditions and significance brackets on the right. Statistical significance is denoted by * ($p < 0.05$), ** ($p < 0.01$), and *** ($p < 0.001$).}
  \label{fig:us1_likert}
\end{figure*}

\paragraph{Task Completion Quality.}

Overall, we found significant main effects of Method ($z = 4.40$, $p < 0.001$) and Task ($\chi^2(3) = 27.64$, $p < 0.001$), and no significant Method $\times$ Task interaction was found ($\chi^2(3) = 1.48$, $p = 0.688$), indicating that ICON's improvement on task completion quality was consistent across tasks. Our analysis further revealed that ICON significantly improved completion quality when aggregating across all tasks ($\Delta = +11.09\%$, $z = 4.40$, $p < 0.001$). Specifically, Holm-adjusted post-hoc tests revealed a similar pattern across tasks, with significant improvements in Pressure Dressing ($\Delta = +13.47\%$, $z = 2.49$, $p = 0.038$) and Pinwheels ($\Delta = +14.35\%$, $z = 2.98$, $p = 0.012$), and numerically consistent but non-significant improvements in Tourniquet ($\Delta = +11.23\%$, $z = 1.98$, $p = 0.095$) and Mug Cake ($\Delta = +5.99\%$, $z = 1.34$, $p = 0.180$).

\paragraph{Task Completion Time.}

We found significant main effects of Method ($z = -2.39$, $p = 0.017$) and Task ($\chi^2(3) = 83.31$, $p < 0.001$), with no significant Method $\times$ Task interaction ($\chi^2(3) = 4.64$, $p = 0.200$). Our analysis further revealed that ICON significantly reduced task completion time when aggregating across all tasks ($\Delta = -15.52\%$, $z = -2.39$, $p = 0.017$). As shown in Fig.~\ref{fig:us1_obj}, post-hoc comparisons showed reductions in three of the four tasks, although no individual task reached statistical significance after Holm correction: Tourniquet ($\Delta = -38.67\%$, $z = -2.24$, $p = 0.092$), Pressure Dressing ($\Delta = -9.02\%$, $z = -0.35$, $p = 1.000$), Pinwheels ($\Delta = +1.03\%$, $z = 0.09$, $p = 1.000$), and Mug Cake ($\Delta = -18.36\%$, $z = -2.28$, $p = 0.092$).

One possible interpretation of why ICON significantly reduced task completion time is that ICON videos were typically shorter than BAU videos. However, in the Mug Cake task, even though the ICON video was substantially longer ($\Delta = +284\%$, ICON: 614 s vs. BAU: 160 s), participants still completed the task faster with ICON videos, suggesting that video length alone could not explain the reduction in task completion time. Furthermore, a robustness analysis controlling for task and
nominal video duration showed that ICON remained associated with approximately 13.6\% shorter completion times ($\beta = -0.147$, $p = .034$; see Supplementary Material, Table~S1).

\paragraph{Cognitive Load.}

The NASA-TLX results (Fig.~\ref{fig:us1_obj}) indicated that participants experienced lower overall workload with ICON than with BAU. Holm-adjusted post-hoc tests revealed that ICON showed significantly lower mental demand ($\Delta = -21\%$, $z = -2.42$, $p = 0.047$), temporal demand ($\Delta = -29\%$, $z = -3.62$, $p = 0.002$), and frustration ($\Delta = -38\%$, $z = -3.10$, $p = 0.008$), and significantly lower scores on the reverse-coded performance subscale, where a lower score indicates better self-rated performance ($\Delta = -32\%$, $z = -3.50$, $p = 0.002$). No significant differences were observed for physical demand ($p = 0.828$) or effort ($p = 0.828$). Overall, ICON showed a significant reduction in the unweighted NASA-TLX score, calculated as the mean of the six subscales after reverse-coding Performance and rescaling to 0--100 ($\Delta = -21.1\%$, $z = -3.10$, $p = 0.002$).

\paragraph{User Perceptions and Preferences.}
Holm-adjusted comparisons on Likert scale responses (Fig.~\ref{fig:us1_likert}) indicated that participants rated ICON videos as significantly easier to understand ($z = 5.612 $, $p < 0.001$), more helpful in saving time on following instructions ($z = 8.443$, $p < 0.001$), and more effective in improving task performance quality ($z = 9.052$, $p < 0.001$) across all tasks. Participants also reported significantly higher satisfaction with their performance after ICON ($z = 2.66$, $p = 0.020$) and higher confidence that they could now complete the task without the video ($p < 0.05$), while ratings of the text instructions' clarity did not differ between conditions (Fig.~\ref{fig:us1_likert}).

Additionally, participants' preferences (Fig.~\ref{fig:us1_obj}, bottom right), analyzed with multinomial logistic models (BAU/ICON/Not Sure), further revealed a significant advantage of ICON over BAU in completion quality ($z = 4.909$, $p < 0.001$), time efficiency ($z = 4.872$, $p < 0.001$), and cognitive load ($z = 4.927$, $p < 0.001$), while preferences for learning effectiveness were more balanced, with no significant difference between ICON and BAU ($z = 0.977$, $p = 0.329$).

Overall, participants' perceptions aligned with their objective performance, suggesting consistency between subjective experiences and observed performance outcomes, with ICON outperforming BAU in task completion quality, task completion time, and cognitive load.

\subsection{Finding 2.1: Identify Key Visual Context Attributes}
Addressing \textbf{RQ2}, we analyzed participants' interview transcripts using affinity diagramming. One researcher first familiarized themselves with the transcripts, tagged comments related to visual matches or mismatches between the video and the real task environment, developed descriptive codes, and grouped related codes into higher-level clusters. A second researcher then reviewed the resulting clusters against the original transcript excerpts, and the authors discussed category boundaries until the four attributes were clearly distinguished. We categorized these into four key visual context attributes: \textit{Task Object Intrinsic}, \textit{Task Object State}, \textit{Environmental Context}, and \textit{Observational Context}.

\paragraph{Task Object Intrinsic.} This attribute includes the intrinsic properties of task-specific objects that participants interact with, such as tourniquets, mannequins, ingredients, and tools. These intrinsic properties typically encompass geometry, color, and texture. In general, most participants ($N = 15$) mentioned that keeping the items the same between the videos and the real world was helpful, since it assisted them in recognizing items and taught them how to use tools correctly: \user{The same ingredients helped me recognize which is salt and which is sugar} (P8), and \user{the video used the same microwave, which told me how to operate it} (P10). Consequently, they did not need to \user{guess or match things up if the video uses the same items} (P8). However, a subset of participants (P5, P7, P10) believed that such consistency was useful but not necessary, since they could still \user{\ldots rely on [themselves] to make up for that} (P7). Furthermore, participants had different perspectives on which items should be kept the same. Although many participants ($N = 10$) preferred to keep only key task-specific items consistent, they differed in how they defined these items, possibly due to varying levels of familiarity. For instance, in the Mug Cake task, some participants (P6, P8) considered spoons to be universal tools that everyone is familiar with, while others (P10, P12, P16) reported that the video helped them identify the spoons and know which one to use. 

\paragraph{Task Object State.} This attribute also involves task-specific objects but focuses on their initial state. For example, objects can be positioned differently, affecting the layout of all task objects. Additionally, the same object can exist in various states, such as a tourniquet being in its bag, laid out, or stretched and wrapped. Participants believed that the inconsistency of task object state had a minor impact on the completion quality: \user{Keeping the location of all the objects the same is not necessary but nice to have} (P4), and in the Pressure Dressing task, \user{If you move the wound to the hand, I can also achieve the same level of completion} (P13). However, many participants mentioned that keeping task objects in the same state could potentially reduce mental effort and improve task completion speed. As P9 noted, \user{In the [BAU] video, the wound is on the right leg, but what I am going to operate on is the left leg. That caused a little bit of difficulty for me,} and P15 explained that \user{having exactly the same location helped with the speed of task completion, since I could see where it is and instantly go to that spot.}

\paragraph{Environmental Context.} This attribute encompasses visual elements not specific to the task but present in the video, often considered as ``background.'' Examples include the table, background objects not required by participants, and lighting conditions. Many participants believed that maintaining a consistent environmental context in the video could help users concentrate on the task. In contrast, excessive visual elements could disrupt focus: \user{A lot of aesthetic stuff [in the BAU video]\ldots was kind of distracting when I was following it. I really like the [ICON] video. It is very simple and straightforward} (P10). Conversely, some participants (P1, P7) claimed that the different background in BAU videos did not distract them from completing the task, as P7 explained: \user{I focus on what's the next step. I don't care what the surrounding environment is.} For the same reason, some participants also proposed that the same environmental context in videos might not have any effect on their task performance.

\paragraph{Observational Context.} Beyond the physical elements, this attribute addresses how the task demonstration is captured and presented in the video. It includes aspects such as filming perspectives, camera movements, and shot composition as highlighted by participants. Most participants noted the observational context could affect both task completion quality and time. First, participants found that different filming perspectives might cause occlusion. For instance, in the Tourniquet task, P11 commented that \user{I have to look more carefully into the [BAU] video because there are occlusions in it,} and P9 favored the ICON videos since \user{the hands did not block me when doing the first step.} Such occlusion might cause confusion and thus require them to rewatch the video and double-check, suggesting a potential impact on task completion time, quality, and cognitive load. Second, participants highlighted that the first-person view in the ICON videos helped them \user{comprehend the task with less time} (P12), because \user{the view of the [ICON] video is the same as my view,} making it \user{easy to understand and follow} (P11). Additionally, P14 proposed that \user{some tutorials have frames from different angles to perform the task,} which could present finer details for complex steps and ultimately enhance the overall understandability of the task.

\label{sec:us2_videos}
\begin{figure}
  \centering
  \includegraphics[width=\linewidth]{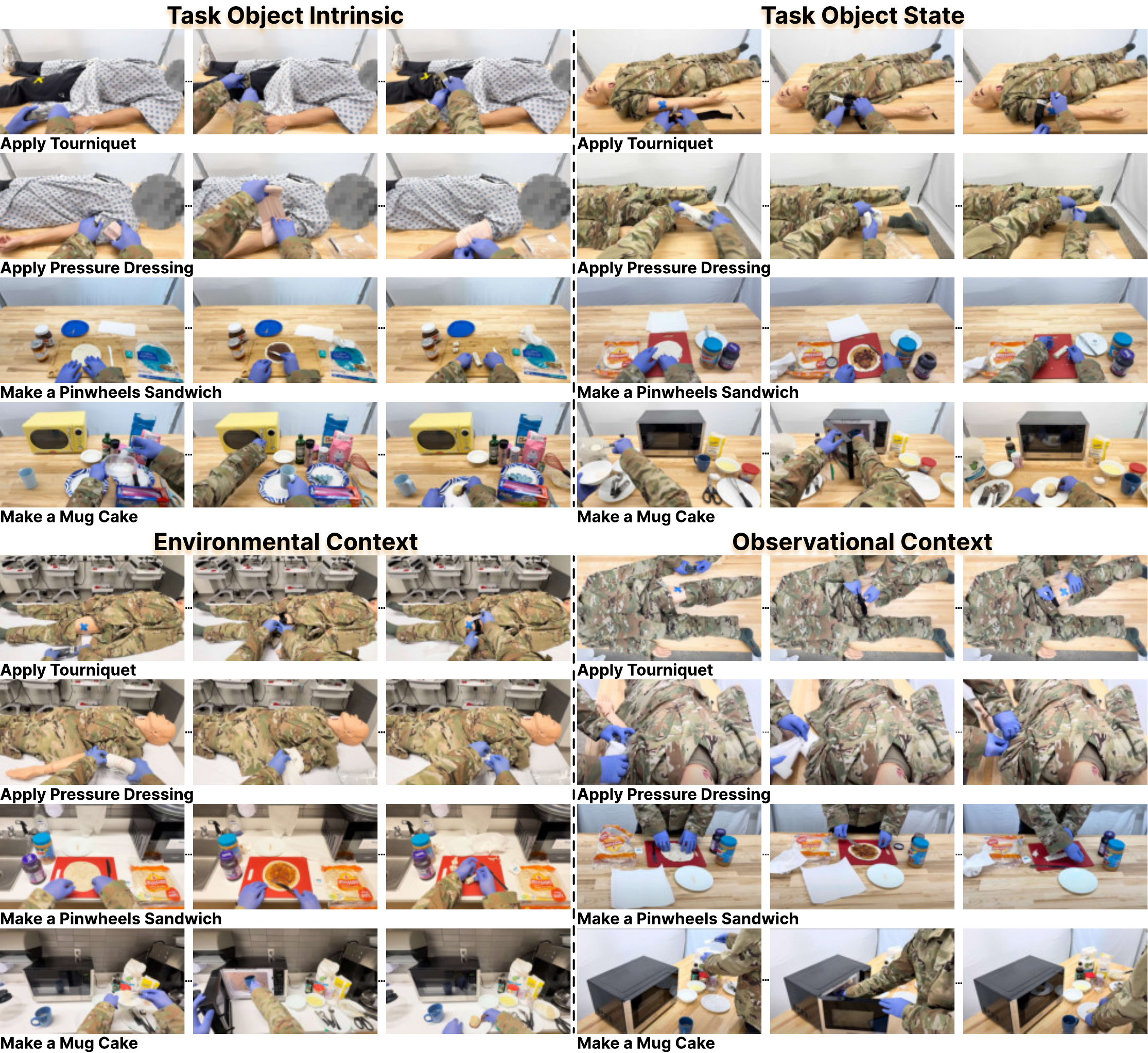}
  \caption{Examples of ablated ICON videos in which we deliberately misalign factors along one and only one of the context attributes at a time.  For example, notice how in the Environmental Context (bottom-left) examples, the backgrounds are different, but the supplies, objects, clothing, and other features are the same as the standard ICON videos.}
  \label{fig:ablated_videos}
\vspace{-15pt}
\end{figure}

\section{Study 2: Measuring the Individual Contribution of Each Context Attribute}
\label{sec:us2}

In Study~1, we concluded a list of key context attributes, while the degree to which each contributes to ICON's effectiveness remains unclear. Study~2 measures each attribute's contribution on its own, pooled across the four tasks, and compares the objective cost of each misalignment with how participants rate the video and their own performance (\textbf{RQ2} and \textbf{RQ3}).

Therefore, we conduct an ablation study on the four context attributes in Study~2, using four videos each with one context misaligned, and an ICON video, resulting in a total of five conditions for comparison. Misaligning one attribute at a time from an otherwise aligned video follows the ablation strategy of Wu and Quinn~\cite{wu2017confusing}, who degraded a crowdsourcing task's instructions one guideline at a time and measured the effect on accuracy, throughput, and satisfaction. We first introduce how we prepared the ablated videos, then describe the experiment design, and finally share the results and findings.

\begin{table*}[t]
\captionsetup{position=bottom}
\centering
\begingroup
\small
\setlength{\tabcolsep}{5pt}
\renewcommand{\arraystretch}{1.25}
\begin{tabularx}{\textwidth}{@{}>{\raggedright\arraybackslash}p{.38\textwidth}*{4}{>{\centering\arraybackslash}X}@{}}
\toprule
\textbf{Condition} & \textbf{Tourniquet} & \shortstack{\textbf{Pressure}\\\textbf{Dressing}} & \textbf{Pinwheels} & \shortstack{\textbf{Mug}\\\textbf{Cake}} \\
\midrule
ICON & 51 & 53 & 241 & 331 \\
Task Object Intrinsic ablation & 49 & 50 & 240 & 330 \\
Task Object State ablation & 38 & 55 & 265 & 326 \\
Environmental Context ablation & 50 & 50 & 240 & 332 \\
Observational Context ablation & 47 & 53 & 250 & 360 \\
\bottomrule
\end{tabularx}
\endgroup
\caption{Video durations in Study 2, in seconds, for each condition and task.}
\label{tab:us2-video-durations}
\end{table*}

\subsection{Ablated In-Context Instructional Videos}

\begin{figure}[t]
    \centering
    \includegraphics[width=\linewidth]{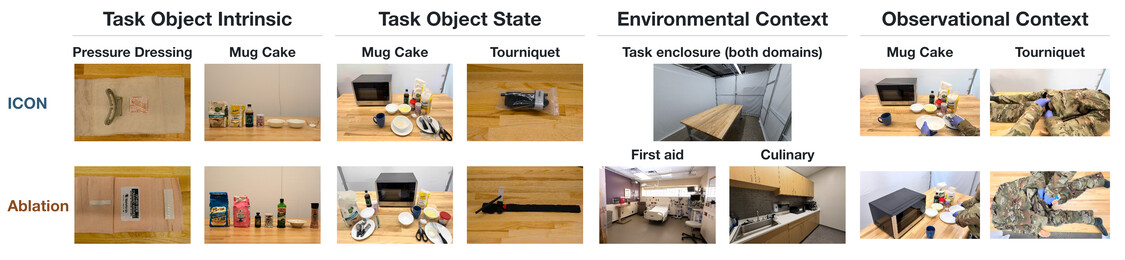}
    \caption{Examples of the four visual-context ablations in Study 2, with ICON shown above and the corresponding ablation below. \textbf{Task Object Intrinsic}: Pressure dressing models differ in color and visible components; the same mug-cake ingredients are presented in different packaging and containers. \textbf{Task Object State}: The same mug-cake tools and ingredients in different positions on the table; the tourniquet is packaged and folded in ICON but unpackaged and extended in the ablation. \textbf{Environmental Context}: In ablation videos, the white-curtained task enclosure is replaced by a clinical setting for first-aid tasks or a kitchen setting for culinary tasks. \textbf{Observational Context}: The first-person ICON view is replaced by a side view for the mug-cake task and an overhead view for the tourniquet task.}
    \label{fig:us2_visual_attribute_comparison}
    \Description{Four panels arranged in one row compare ICON with the corresponding ablations. Task Object Intrinsic shows different pressure dressing models and ingredient containers. Task Object State shows rearranged mug-cake supplies and a packaged versus unpackaged tourniquet. Environmental Context compares the curtained enclosure with clinical and kitchen settings. Observational Context compares first-person views with side and overhead views.}
\end{figure}

We illustrate the ablated videos key frames (showing starting, middle and end stage of the tasks) in Fig.~\ref{fig:ablated_videos}. Our ablations were designed as natural alternatives. Each ablated video remained a plausible instructional video while differing from ICON along one target visual context attribute. All conditions shared the same task goal, step sequence, instructional content, and demonstration personnel. Video durations were not identical (Table~\ref{tab:us2-video-durations}); the largest gap is the Task Object State video for the tourniquet, 38 s against ICON's 51 s, because the tourniquet starts unpackaged and extended in that condition (Fig.~\ref{fig:us2_visual_attribute_comparison}), so the unpacking shown in the other conditions does not appear.
We shortened the videos for Study 2 in response to Study 1's interview feedback. Participants described ICON as slow, suggested a quicker walkthrough, and noted that inactive waiting periods, such as a one-minute microwave cycle, could be skipped. ICON's definition remained unchanged: the video matches the participant's task objects, object states, environment, and viewing perspective. Because video timing and the study design differed between Study 1 and Study 2, we interpret comparisons within each study and do not directly compare performance across studies.

Study 2's ablated conditions were prepared relative to this revised ICON reference (Table~\ref{tab:us2-ablation-comparison}). When misaligning a given context attribute, we ensured that the ablated video is still a plausible and helpful instructional video with just the targeted attribute differing. 

Specifically, in the \textbf{Task Object Intrinsic} ablation, we changed stable visual properties of task-relevant objects, such as object model, geometry, color, material, or texture. For example, first-aid videos used a different tourniquet or bandage, while culinary videos used different ingredient containers, plates, utensils, cutting boards, or microwave models and colors. (Figure~\ref{fig:us2_visual_attribute_comparison}, \textit{Task Object Intrinsic} panel)

In the \textbf{Task Object State} ablation, we kept object identities largely the same but changed their configuration, placement, orientation, or preparation state, such as the position of the wound, the arrangement of supplies on the
table, or the preparation stage and location of ingredients and tools. (Figure~\ref{fig:us2_visual_attribute_comparison}, \textit{Task Object State} panel)

In the \textbf{Environmental Context} ablation, we preserved the task objects and demonstration but changed the surrounding non-task context. For example, first-aid tasks were shown in an emergency-room-like setting instead of the experimental
task area, while culinary tasks were shown on a kitchen counter with different background appliances, surfaces, and lighting. (Figure~\ref{fig:us2_visual_attribute_comparison}, \textit{Environmental Context} panel)

In the \textbf{Observational Context} ablation, we preserved the physical setup but changed how the task was filmed, such as replacing the egocentric ICON view with third-person, side, overhead, or more occluded viewpoints. (Figure~\ref{fig:us2_visual_attribute_comparison}, \textit{Observational Context} panel)

\begin{table*}[t]
\captionsetup{position=bottom}
\centering
\begingroup
\small
\setlength{\tabcolsep}{4pt}
\renewcommand{\arraystretch}{1.15}
\begin{tabularx}{\textwidth}{@{}>{\raggedright\arraybackslash}p{.20\textwidth}*{4}{>{\centering\arraybackslash}p{.091\textwidth}}>{\raggedright\arraybackslash}X@{}}
\toprule
\textbf{Condition} & \multicolumn{4}{c}{\textbf{Visual context}} & \textbf{Explanation} \\
\cmidrule(lr){2-5}
& \shortstack{Task Object\\intrinsic} & \shortstack{Task Object\\state} & \shortstack{Environ.\\context} & \shortstack{Observ.\\context} & \\
\midrule
\textbf{ICON} & \multicolumn{4}{c}{Reference video} & Task-matched objects, states, and environment; fixed first-person view. \\
\addlinespace[5pt]
Ablation: Task Object Intrinsic & \ICONRevChangedCell & -- & -- & -- & Different object appearances (e.g., mannequin appearance/clothing; containers, tools, ingredients, or microwave). \\
\addlinespace[5pt]
Ablation: Task Object State & -- & \ICONRevChangedCell & -- & -- & Different object placement, orientation, or preparation state (e.g., wound position or supply layout). \\
\addlinespace[5pt]
Ablation: Environmental Context & -- & -- & \ICONRevChangedCell & -- & Different non-task surroundings (e.g., an emergency-room-like setting or kitchen background). \\
\addlinespace[5pt]
Ablation: Observational Context & -- & -- & -- & \ICONRevChangedCell & Different viewpoint, framing, or occlusion (e.g., third-person, side, or overhead view). \\
\bottomrule
\end{tabularx}
\endgroup
\caption{Study 2 video conditions. ICON matches all four visual-context attributes. Each ablation changes one attribute while retaining the other three. Illustrations of each condition are shown in Fig.~\ref{fig:us2_visual_attribute_comparison}. Task goals, step sequence, instructional content, and narration are shared across conditions. Video durations are not identical; values for each task and condition are reported in Table~\ref{tab:us2-video-durations}.}
\label{tab:us2-ablation-comparison}
\end{table*}

\subsection{Study Design}
We conduct a between-subjects study with five Context 
Condition levels: full alignment (ICON), misaligned task object intrinsic, misaligned task object state, misaligned environmental context, and misaligned observational context. We recruited 40 participants (8 per condition), each completing all four tasks once under a single assigned condition. After each task, 
participants complete a NASA-TLX survey and a 7-point Likert scale assessment, followed by a semi-structured interview covering their overall experience, challenges encountered, and perceived differences between the video and their actual setup (question list in Supplementary Section~S3).

With $N{=}40$ ($n{=}8$/cell), a two-sided per-contrast test at $\alpha = 0.05$ has 80\% power to detect large differences (Cohen's $d \approx 1.5$), so the study is powered for large effects and smaller ones may not reach significance.
To counterbalance task-ordering effects (e.g., fatigue), the eight participants in each condition follow four cyclic task orders, two participants per order (a partial Latin square).
As a result, each participant completes a 1 to 1.5-hour session, compensated \$40, yielding 51 hours of task performance data in total. Participant identifiers in this section restart at P1 and are independent of Study~1's.

\subsection{Finding 2.2: Removing Visual Context Objectively Degrades Task Performance}
\begin{figure*}[t]
\centering
\includegraphics[width=\linewidth]{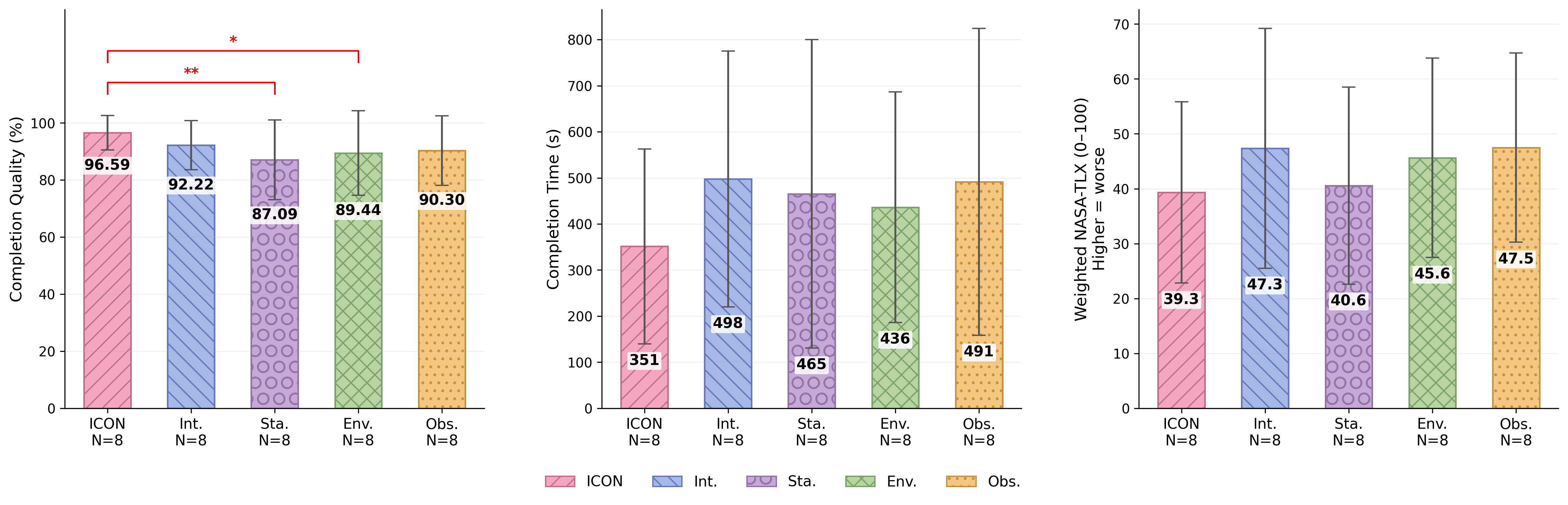}

\caption{Ablation study results comparing ICON with four ablated conditions (Int.\ = Task Object Intrinsic, Sta.\ = Task Object State, Env.\ = Environmental Context, Obs.\ = Observational Context).
\textbf{Left:} Completion quality (\%).
\textbf{Center:} Completion time (s).
\textbf{Right:} Weighted NASA-TLX scores (0--100).
Statistical significance is denoted by * ($p < 0.05$)
and ** ($p < 0.01$).}
\label{fig:ablation_objective}

\par\vspace{10pt}

\begin{minipage}{\linewidth}
    \centering
    \begingroup
    \small
    \setlength{\tabcolsep}{5pt}
    \renewcommand{\arraystretch}{1.2}

    \begin{tabular*}{\linewidth}
        {@{\extracolsep{\fill}}lcccccc@{}}
        \toprule
        & \multicolumn{3}{c}{\textbf{Main Effect of Method}}
        & \multicolumn{3}{c}{\textbf{Method $\times$ Task Interaction}} \\
        \cmidrule(lr){2-4}
        \cmidrule(lr){5-7}

        \textbf{Metric}
        & $\chi^2$ & \textit{df} & \textit{p}
        & $\chi^2$ & \textit{df} & \textit{p} \\
        \midrule

        Completion Quality (\%) $\uparrow$
        & 11.50 & 4 & \textbf{0.022}\textsuperscript{*}
        & 24.06 & 12 & \textbf{0.020}\textsuperscript{*} \\

        Completion Time (s) $\downarrow$
        & 8.19 & 4 & 0.085
        & 14.35 & 12 & 0.279 \\

        Weighted NASA-TLX $\downarrow$
        & 4.95 & 4 & 0.292
        & 8.02 & 12 & 0.783 \\

        \bottomrule
    \end{tabular*}
    \endgroup

    \captionof{table}{Linear mixed model (LMM) analysis
    results for Study 2. One model was fitted per metric:
    \texttt{Metric $\sim$ C(Method, Treatment(ICON)) *
    C(Task) + (1 | Participant)}.
    Both sets of columns report Wald $\chi^2$ tests from
    this single interaction model.
    The \textit{main effect of Method} is the joint Wald
    $\chi^2$ test (df = 4) that all four average marginal
    simple effects of each ablated condition relative
    to ICON are zero, with equal weight across the four tasks.
    The \textit{Method $\times$ Task} columns report the
    joint Wald $\chi^2$ test (df = 12) over all interaction
    coefficients. For NASA-TLX, Performance was reverse-scored before applying the pairwise-comparison weights and rescaling to 0--100, so that higher scores indicate greater workload. Significant p-values are bolded;
    \textsuperscript{*}$p < 0.05$,
    \textsuperscript{**}$p < 0.01$,
    \textsuperscript{***}$p < 0.001$.}
    \label{tab:us2_lmm_omnibus}
\end{minipage}

\end{figure*}

To examine how different components of ICON contribute to task completion quality and completion time, we used the same linear mixed-effects model as in Study 1, with \textit{Method} (ICON and ablated variants), \textit{Task}, and their interaction (\textit{Method} $\times$ \textit{Task}) as fixed effects, and a random intercept for each participant. Post-hoc pairwise comparisons (ICON vs.\ each ablated condition) were Holm-adjusted. Table~\ref{tab:us2_lmm_omnibus} reports the omnibus tests for each metric. \textbf{Overall, we observed a trend that removing any visual attribute can negatively impact task completion quality and task completion time when pooled across tasks.} We explain the findings in two separate sections below.

\paragraph{Completion Quality.}
We found a significant main effect of \textit{Method} ($\chi^2(4) = 11.50$, $p = 0.022$) and a significant \textit{Method} $\times$ \textit{Task} interaction ($\chi^2(12) = 24.06$, $p = 0.020$), suggesting that the contribution of each visual attribute varies across tasks. In pooled all-task comparisons, ICON achieved higher quality than all four ablated conditions (Fig.~\ref{fig:ablation_objective}, left). Specifically, significant reductions were observed for Environmental Context ($\Delta = -7.40\%$, $z = -2.40$, $p = 0.049$) and Task Object State ($\Delta = -9.83\%$, $z = -3.19$, $p = 0.006$), while Task Object Intrinsic ($\Delta = -4.52\%$, $z = -1.47$, $p = 0.142$) and Observational Context ($\Delta = -6.51\%$, $z = -2.11$, $p = 0.069$) showed non-significant but consistent decreases.

\paragraph{Completion Time.}
We found a marginal main effect of \textit{Method} ($\chi^2(4) = 8.19$, $p = 0.085$) and no significant \textit{Method} $\times$ \textit{Task} interaction ($\chi^2(12) = 14.35$, $p = 0.279$), suggesting that the impact of removing attributes on completion time is relatively consistent across tasks. In pooled all-task comparisons (Fig.~\ref{fig:ablation_objective}, center), removing each attribute increased completion time relative to ICON: Task Object Intrinsic ($\Delta = +41.68\%$, $z = 2.49$, $p = 0.051$), Observational Context ($\Delta = +39.84\%$, $z = 2.38$, $p = 0.052$), Task Object State ($\Delta = +32.45\%$, $z = 1.94$, $p = 0.105$), and Environmental Context ($\Delta = +24.19\%$, $z = 1.45$, $p = 0.148$). Although none of these differences reached statistical significance after Holm correction, they showed consistent directional increases in completion time when each attribute was removed.

\begin{figure}[t]
  \centering
  \includegraphics[width=\columnwidth]{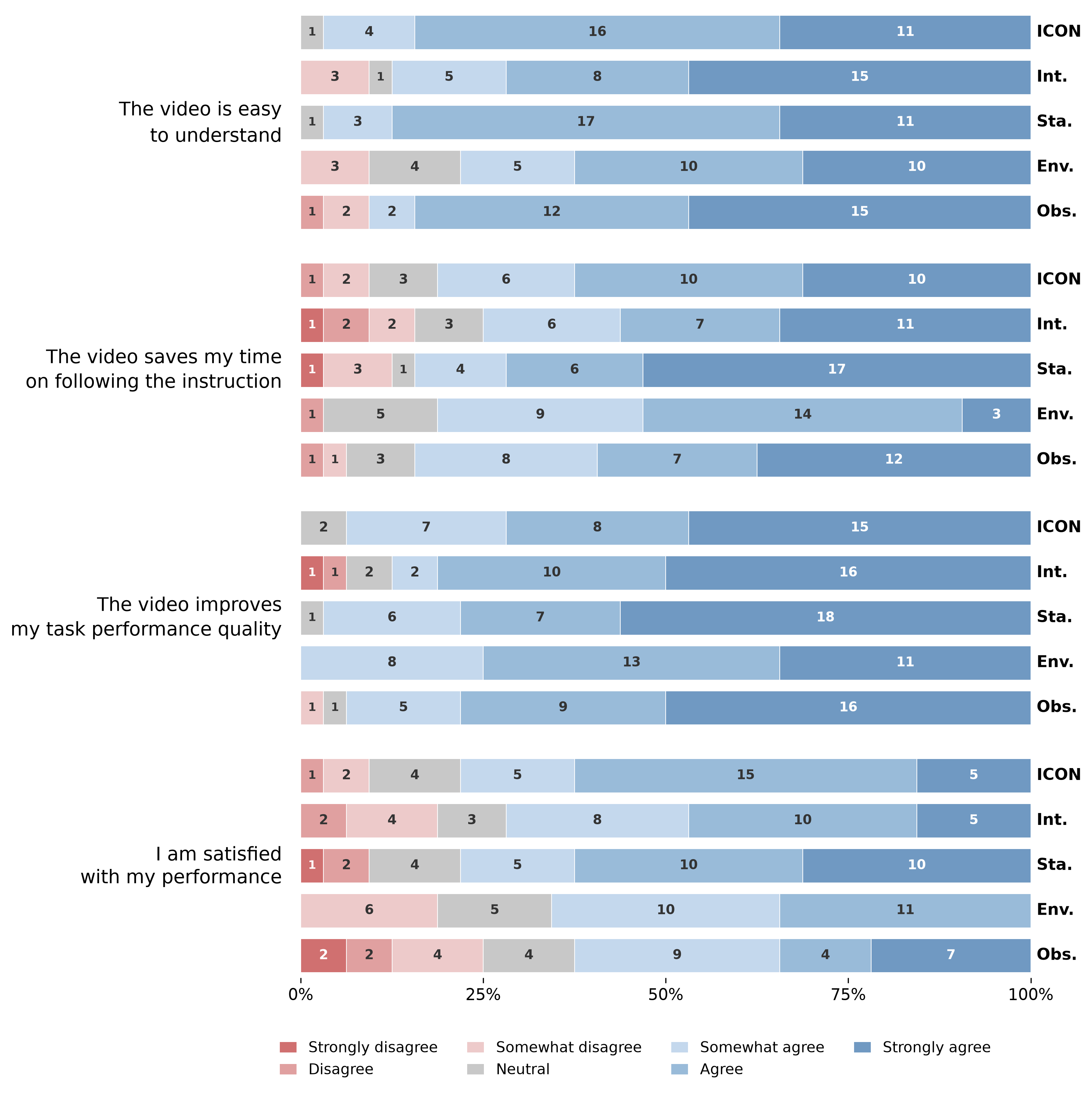}
  \caption{Likert-scale responses from the ablation study comparing ICON with four ablated conditions (Int.\ = Task Object Intrinsic, Sta.\ = Task Object State, Env.\ = Environmental Context, Obs.\ = Observational Context) across four subjective measures: video understandability, perceived time savings, perceived quality improvement, and satisfaction with performance.}
  \label{fig:ablation_subjective}
  \vspace{-10pt}
\end{figure}

\subsection{Finding 3.1: Users Do Not Notice the Objective Cost}
Across all five conditions, participants rated their instructional videos positively. %
\textbf{Yet these ratings revealed no significant differences between ICON and any ablation condition}: Holm-adjusted post-hoc tests on video understandability, perceived time savings, perceived completion quality improvement, and satisfaction with performance all yielded $p > 0.1$ (Fig.~\ref{fig:ablation_subjective}). %
Regardless of which video they watched, participants converged on the same level of self-perceived performance, even those whose objective completion quality dropped by up to 9.83\% and whose completion time increased by up to 41.68\% relative to ICON (Finding~2.2). The measured ranking also inverts what participants had anticipated in Study 1, where they judged Task Object State a minor factor and Task Object Intrinsic an important one (Finding 2.1): in Study 2, object state produced the largest quality loss and object appearance the smallest, although the latter produced the largest numerical increase in completion time. Weighted NASA-TLX scores did not differ significantly between ICON and any ablation condition (Finding 3.2). %
Cognitive science predicts exactly this outcome. When material feels easy to process, people treat that fluency as a signal that they are performing well~\cite{oppenheimer2008metacognition}. Our participants behaved accordingly: even in ablated conditions, they described the videos as sufficient. \user{At least I can see everything needed in the video clearly} (P3, Observational Context). \user{[The video] conveys all the information that I need, which is good enough} (P4, Task Object State). The near-aligned videos retained enough contextual cues to feel fluent, and that fluency became participants' proxy for quality. %
This fluency-driven confidence is what Koriat and Bjork~\cite{koriat2005illusions} call an ``illusion of competence'': conditions that feel fluent produce overconfidence that persists even under challenge. We observed this directly. When we disclosed the manipulation at the end of the session and named the attribute that had been misaligned, P7 (Environmental Context condition) responded, \user{Oh, I didn't even notice that.} P9 (Task Object Intrinsic condition), told that the tourniquet model in the video differed from theirs, said, \user{Of course. I just took it as the same thing.} Most strikingly, even after these disclosures, participants held firm: \user{The video is showing exactly what I should do anyway} (P11, Task Object State). The illusion did not break when confronted with evidence. %
Change blindness compounds the effect. Observers routinely miss visual changes outside their attentional focus~\cite{simons1999gorillas}, and when following an instructional video, attention is locked on procedural content (e.g., what to do next, which tool to pick up), not on whether the depicted objects and surroundings match reality. When we asked participants ``How do you think the visual content in the video could be improved?'', several paused for an extended time before conceding they could not think of anything (P5, Observational Context; P14, Environmental Context). They were not simply satisfied; the mismatch was not available to them as a category of improvement even when the question invited it. %

Together, these three mechanisms explain why subjective evaluation is structurally unable to detect the performance costs of visual context misalignment. The practical implication is clear: relying on user self-reports alone would lead designers to conclude that context alignment does not matter, when in fact it produces measurable gains that users simply cannot perceive. %

\subsection{Finding 3.2: All Ablations Show Numerically Higher Overall Workload}

Overall workload, measured by the weighted NASA-TLX score, was numerically higher in all four ablation conditions than in ICON, although none of these differences was statistically significant. \textit{Method} had no significant main effect ($\chi^2(4) = 4.95$, $p = 0.292$), and the interaction with \textit{Task} was also nonsignificant ($\chi^2(12) = 8.02$, $p = 0.783$). In the pooled Holm-adjusted post-hoc comparisons, the largest numerical increase relative to ICON was observed for Observational Context ($\Delta = +8.15$, $z = 1.661$, $p = 0.386$), closely followed by Task Object Intrinsic ($\Delta = +8.01$, $z = 1.634$, $p = 0.386$), then Environmental Context ($\Delta = +6.29$, $z = 1.283$, $p = 0.399$), and Task Object State ($\Delta = +1.23$, $z = 0.251$, $p = 0.802$; Fig.~\ref{fig:ablation_objective}, right). These differences are expressed on the 0--100 scale, and their ordering is descriptive rather than evidence of significant differences between ablations. None of the six NASA-TLX dimensions differed significantly between ICON and any ablation on its own.

This numerical pattern is compatible with Finding~3.1. Participants in all four ablation conditions reported higher overall workload on average, yet their ratings of their performance and the video did not differ significantly from those of the ICON group. A possible interpretation is that visual context misalignment requires additional effort without participants interpreting that effort as a sign that the video is hurting their outcome.

\section{Discussion, Limitations, and Future Work}
\label{sec:discussion}

\begin{figure}
  \centering
  \includegraphics[width=\linewidth]{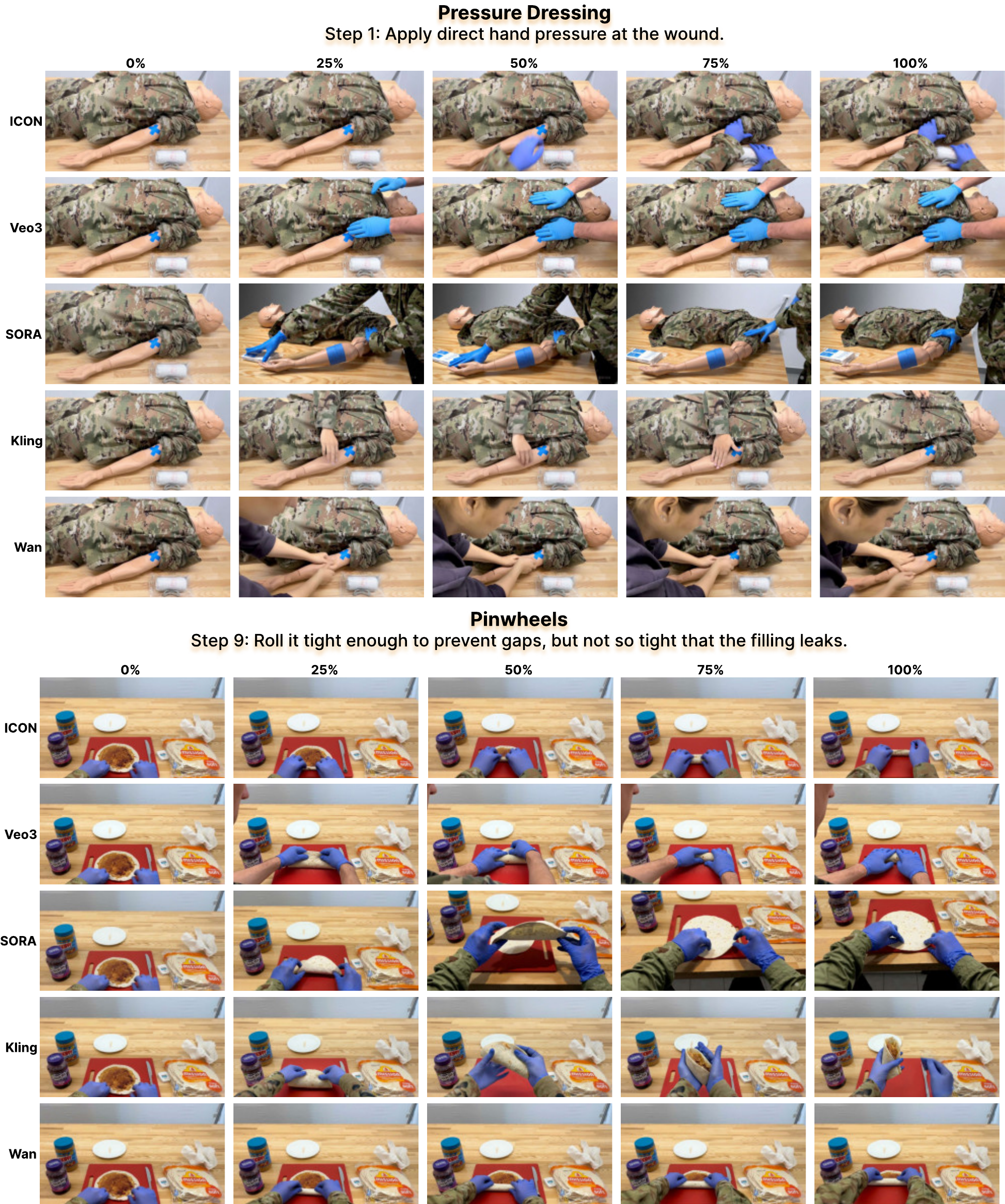}
  \caption{
  Comparative visual examples of state of the art TI2V generation techniques for two steps of two tasks (arbitrarily chosen: the challenges observed here occur frequently across steps).  Although the results are promising, there are clear physical impossibilities shown from each method, demonstrating a need for more work in instructional video generation to support scaling ICON.
  }
  \label{fig:vidgen}
\vspace{-15pt}
\end{figure}

\subsection{Assessing the Scalability of In-Context Instructional Videos}
\label{sec:vidgen}

With the effectiveness of ICON over BAU validated and the contributive visual context surfaced, a natural next question is how to acquire ICON at scale. It is simply not feasible to manually record ICON for every arbitrary physical task in every arbitrary setting.  With recent advances in AI video generation, e.g., ~\cite{dai2023fine,chen2024videocrafter2,ni2024ti2v,yang2024cogvideox}, we see a potential avenue for automatically generating ICON videos.
Like recent advances in instructional image generation~\cite{lai2023lego, souvcek2024genhowto}, it seems plausible that these video generation advances could be a solution for scaling in-context instructional video generation.

Instructional video generation is captured by the well-known text-and-image-to-video (TI2V) problem. TI2V methods, such as TI2V-Zero~\cite{ni2024ti2v} and PIA~\cite{zhang2024pia}, extend text-to-image (T2I) models~\cite{rombach2022high, saharia2022photorealistic} to produce videos conditioned on images and text.  Although this general problem setting is what we need to scale ICON, the current state of the art is limited.  First, most generation techniques focus on generation of full scenes rather than the focused generation of activities within the scene.  One method that is able to focus on targeted regions is AnimateAnything~\cite{dai2023fine}. Second, contemporary methods focus on visual quality more than physical realism, challenging their utility for scaling ICON, at this point.

Figure~\ref{fig:vidgen} presents a visual case study of these limitations on the physical tasks of this paper.  Here, we choose four contemporary TI2V generation techniques: Google Veo3~\cite{GoogleVeo3_2025}, OpenAI Sora~\cite{OpenAI_Sora_2024_WorldSim}, Kling 2.0~\cite{Kling_2_1_2025}, Wan 2.2~\cite{wan2025wanopenadvancedlargescale}.  For two task-steps, we sample the generated frames at five uniformly distributed frames through the video.  For example, the Sora method in the Pressure Dressing Step 1 completely changes the context.  The Kling method has the mannequin reaching over and applying pressure with its own right hand, and the Veo3 method has pressure being placed in two locations.  Similar challenges are observed in the Pinwheels Step 9 case with the tortilla being folded instead of rolled (Kling and Veo3) or the tortilla losing its filling (Sora).  Read against the four attributes of Sec.~\ref{sec:us2}, three of the four models preserve the environment, the objects' appearance, and their starting arrangement, Sora replaces all three, and Wan lets another person's body enter the frame; the remaining errors, pressure applied away from the wound and a tortilla folded rather than rolled, concern the demonstrated action rather than its visual context, so matching the four attributes is necessary but not sufficient for a usable generated video.

In summary, although AI-based video generation techniques represent a promising direction for scaling the ICON findings in this paper, the current state of the art methods are insufficient and more work is needed.

\subsection{Performance vs. Learning}
\label{sec:implications}

Our studies measured task completion, not learning, but Study~1's preference data touch on it (Fig.~\ref{fig:us1_obj}, bottom right). Participants chose ICON over BAU four times out of five for completion quality, time, and cognitive load, yet for learning the choice was split and not significant. Interviews point to the narration rather than the visuals. P11, who chose BAU for learning, explained: ``I feel I can do the task out of this room.'' The BAU videos carry explanatory content that ICON's text-to-speech reading of the instructions omits, such as ``Importantly, the tourniquet may be placed over clothing.'' At the same time, participants reported more confidence that they could now complete the task without the video after ICON than after BAU (Fig.~\ref{fig:us1_likert}), so the preference data do not show a learning cost of alignment; they show that participants valued BAU's explanations.

Whether visual alignment helps or hinders skill acquisition, and how explanatory content should be combined with it, remains open. A plausible design keeps full visual alignment for the demonstration and adds explanatory content about when variations are acceptable; testing it requires a learning outcome measured after a delay, which our studies did not include. The question is relevant well beyond the lab, given the many ways instructional videos are used in professional training~\cite{morgado2024video}.

\subsection{Threats to Validity}

Our work has several notable limitations related to the evaluation of physical task completion.  First and most obvious, in many (or most) physical tasks, it is plausible to successfully accomplish a task without exactly following the steps in written instructions or a visual aid.  Hence, measuring task outcomes is different than measuring task step following.  In our study, we measured completion time and a step-following quality score, in which experts grade each instructed step on a rubric; the score therefore rewards following the steps rather than the outcome as such.  However, the relationship between task outcome and task steps is an interesting one.  For example, the significant Method $\times$ Task interaction on quality in Study~2 shows that attributes matter differently across tasks, and they plausibly matter differently across steps within a task; we did not analyze effects at the step level.  There is room for follow-on work that more closely examines the relationship between task step completion and task outcome, including the possibility of \textit{rogue} task process modification on the flow by study participants, still leading to a successful outcome.

Second, the task completion metric conflates performance time---the time to actually carry out the physical task once started---with whole task time, which is the combined time it takes to both carry out the task, to read the written description, and to watch the instructional video.  We do not distinguish between these two in this paper.  Third, although our studies span two classes of physical tasks, four individual tasks, and are carefully designed to allow strong analysis, we do not present a mechanism to measure how generalizable these tasks are or how well these tasks cover the space of possible physical manipulation tasks.  Each of these three aspects represents interesting questions for future inquiry.

Three further limitations concern the evidence itself. Study~1 compares Internet videos with fully aligned recordings that also differ in narration, step order, and editing (Table~\ref{tab:us1-video-comparison}); it measures the total gap between the status quo and full alignment, and the attribution of the cost to visual context rests on Study~2, where only one attribute changes per condition. Finding~3 rests on ratings that did not differ and on what participants said when asked; we did not run a manipulation check that measures detection directly, such as a forced-choice task, and with eight participants per condition the absence of a rating difference is an observed null rather than demonstrated equivalence. Finally, the quality rubric was calibrated on two participants' recordings before independent grading.

\section{Conclusion}
We find visual context misalignment in instructional videos is substantial, decomposable into four attributes, and invisible to users. %
The 11\% quality gain and 15.5\% time reduction we measured (F1) give empirical grounding to what motor simulation and cognitive load theories have long predicted: visual context alignment matters for physical task performance. Future work can apply the same study workflow to estimate how much in-context video could help for other physical tasks. %
The four-attribute decomposition (F2) offers an evaluation framework for evaluating AI-generated instructional videos, complementing existing video generation metrics that measure visual fidelity but not instructional usefulness. %
The finding that users do not notice the effect of single-attribute misalignment (F3), even as their performance drops measurably, echoes the ``illusion of competence'' documented in metacognitive monitoring research~\cite{koriat2005illusions}. Its practical consequence is the one to take away: asking viewers whether an instructional video fits them will say yes when it does not, so fit has to be measured. %
Together, these findings help further understand how visual context misalignment in instructional video affects physical task completion.  %

\begin{acks}
This research was funded, in part, by the U.S. Government under ARPA-H contract 1AY2AX000062. The views and conclusions contained in this document are those of the authors and should not be interpreted as representing the official policies, either expressed or implied, of the U.S. Government.
\end{acks}

\bibliographystyle{ACM-Reference-Format}
\bibliography{handi,sample-base}

\clearpage
\appendix
\onecolumn

\section*{Supplementary Material}
\setcounter{section}{0}\setcounter{figure}{0}\setcounter{table}{0}
\renewcommand{\thesection}{S\arabic{section}}\def\theHsection{S\arabic{section}}
\renewcommand{\thefigure}{S\arabic{figure}}\def\theHfigure{S\arabic{figure}}
\renewcommand{\thetable}{S\arabic{table}}\def\theHtable{S\arabic{table}}

This supplementary material accompanies the main paper. Its sections, figures, and tables carry an ``S'' prefix (S1, S2, \ldots) to distinguish them from those in the main paper.

\section{Task Description}
\label{sec:task_text}
\begin{figure}[htbp]
\centering
  \includegraphics[
    width=0.9\linewidth,
    height=0.8\textheight,
    keepaspectratio
  ]{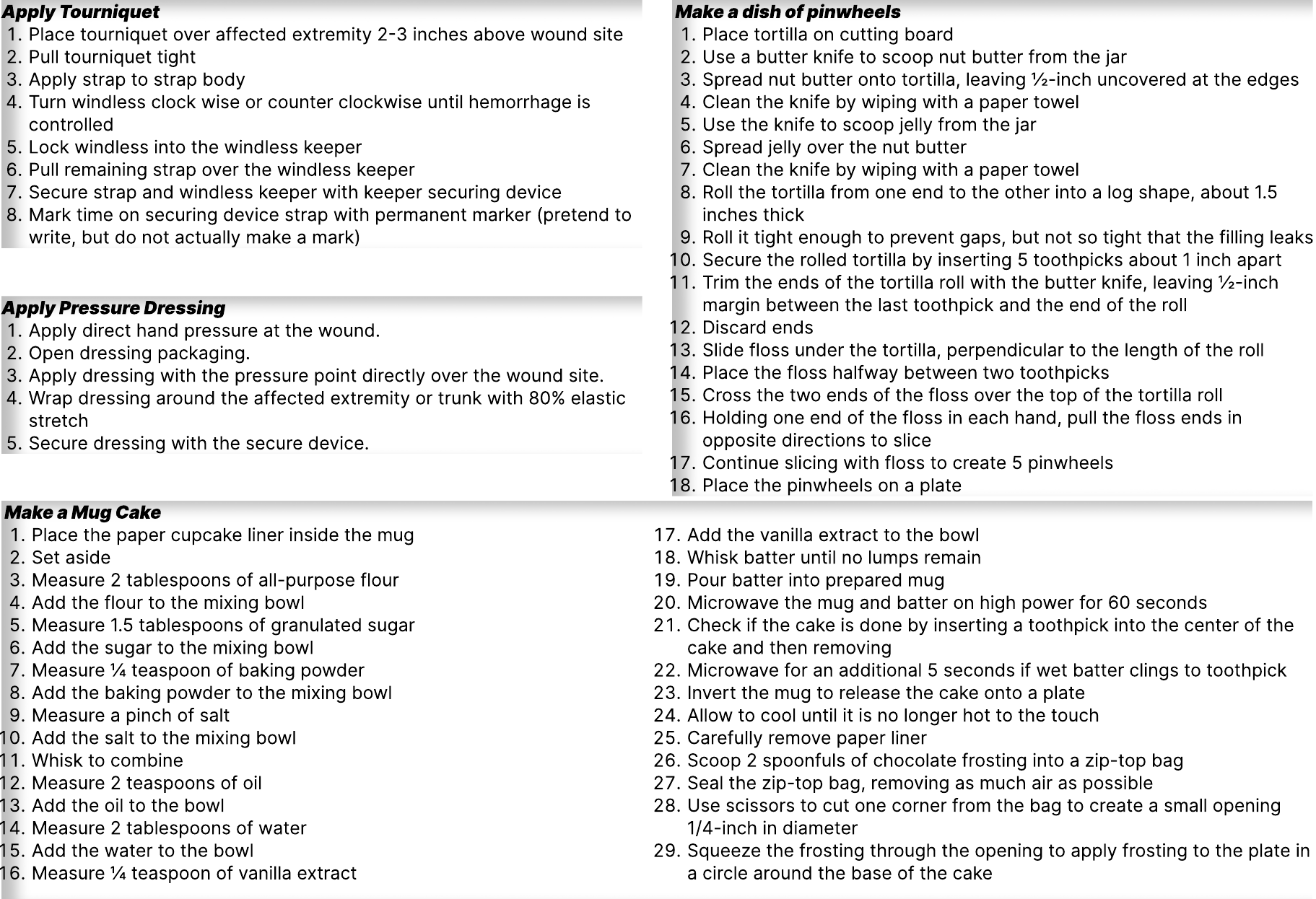}
  \caption{
  The exact textual instructions given to study participants for each of the four physical tasks.
  }
  \Description{Four text panels stacked vertically, one per task, each with a bold title and numbered steps: Apply Tourniquet (8 steps), Apply Pressure Dressing (5 steps), Make a dish of pinwheels (18 steps), and Make a Mug Cake (29 steps).}
  \label{fig:task_text}
\end{figure}

Fig.~\ref{fig:task_text} shows the exact textual description for each of the four physical tasks used in our study.

\clearpage
\section{Completion-Time Robustness Analysis}
\label{sec:time-robustness}

\begin{table}[htbp]
    \centering
    \small
    \renewcommand{\arraystretch}{1.2}
    \setlength{\tabcolsep}{6pt}

    \begin{tabularx}{\linewidth}{
        @{}>{\raggedright\arraybackslash}p{0.32\linewidth}
        >{\raggedright\arraybackslash}X
        r@{}
    }
        \toprule
        \textbf{Effect / comparison} &
        \textbf{Estimate and test statistic} &
        \textbf{$p$-value} \\
        \midrule
        Method (ICON vs.\ BAU) &
        $\beta = -0.147$, SE $= 0.069$, $z = -2.12$ &
        .034 \\

        \addlinespace[4pt]
        Nominal video length (s) &
        $\beta = -0.00009$ &
        .768 \\

        \midrule
        Adding video length &
        Likelihood-ratio $\chi^2(1) = 0.091$ &
        .763 \\
        \bottomrule
    \end{tabularx}

    \caption{Robustness analysis of Study 1 completion time.
    The linear mixed-effects model predicted log-transformed
    completion time from Method, Task, and nominal video length,
    with a random intercept for participant.
    ICON remained associated with approximately 13.6\% lower
    completion time than BAU after controlling for video length.
    The final row compares models with and without video length;
    adding this predictor did not significantly improve model fit.}
    \label{tab:us1-time-robustness}
\end{table}

\section{Interview Protocols}
\label{sec:interviews}
Both studies used semi-structured interviews. The scripts below guided them; the interviewer followed up on answers as needed. Interviews were recorded and transcribed. Bracketed text is an interviewer instruction or a definition offered on request, not a question.

\subsection{Study 1 (within-subjects, BAU vs.\ ICON)}
\textbf{After every trial} (once the post-trial survey is complete).
\begin{enumerate}
  \item Can you describe your experience just now? Is there anything that you want to share?
\end{enumerate}

\textbf{After the second trial of each task} (the participant has now used both video types on the same task; asked after the preference survey, in addition to the question above).
\begin{enumerate}
  \item Have you ever used these tools or performed similar tasks before?
  \item What is the difference in the way you used the video this time compared with the previous time?
  \item Could you elaborate on your choices in the preference survey?
\end{enumerate}

\textbf{Closing interview} (after all eight trials). [Show the participant the videos they used in the study.]
\begin{enumerate}
  \item The [first video type] videos were the first kind of video we provided, and the [second video type] videos were the second kind. In general, what is the difference between these two kinds of video?
  \item Did you experience different levels of cognitive load when watching and performing tasks with these two types of video? [Cognitive load refers to how much mental effort or concentration is required to understand and complete a task.]
  \item Which kind of video can enhance efficiency? Why? Please elaborate on the quality of task completion and the time it takes to complete.
  \item Which kind of video can enhance learning? Why? [Learning is the process of acquiring knowledge and skills.]
  \item In both kinds of video, what components or features are helpful for you? What are the differences between those components, in their functionality and purpose?
  \item If you were invited to give a name to this kind of video, what would you call it? Why?
\end{enumerate}

\subsection{Study 2 (between-subjects, five context conditions)}
\textbf{After every trial} (once the NASA-TLX and the post-trial survey are complete).
\begin{enumerate}
  \item Can you describe your experience just now?
  \item Do you think the visual content in the video is helpful? In what way?
  \item How do you think the visual content in the video could be improved? [Prompt: think about the challenges you had.]
  \item Did you notice any visual difference between the video and your actual setup? Did it affect your cognitive load, or the completion time and quality of this task?
\end{enumerate}

\textbf{Closing interview} (after all four trials). In the four misaligned conditions, the interviewer first discloses the manipulation: ``By design, we kept most of the visual content in the video aligned with your setting. However, we intentionally kept one factor out: [the attribute misaligned in the participant's condition].''
\begin{enumerate}
  \item Did you notice the difference?
  \item What do you think is the general effect of the misalignment?
  \item {[Going through the task videos.]} Does this one-factor misalignment have different effects on different tasks? How?
  \item Do you think it would be better if the video showed [the option matching the participant's condition: a first-person view (Observational Context), the same appearance (Task Object Intrinsic), the same object states and layout (Task Object State), or the same environment (Environmental Context)]? How would the difference be?
\end{enumerate}

\end{document}